\documentclass[fleqn,usenatbib]{mnras}
\usepackage{newtxtext,newtxmath}
\usepackage[T1]{fontenc}
\usepackage{ae,aecompl}
\usepackage{color,soul}

\PassOptionsToPackage{normalem}{ulem}
\usepackage{ulem}

\newcommand{\added}[1]{#1}
\newcommand{\deleted}[1]{}

\usepackage{graphicx}	
\usepackage{wasysym}

\newcommand{\angstrom}{\textup{\AA}}

\title[SNe Ia head-on collision model polarization] {Polarization signatures of the head-on collision model for Type Ia supernovae: How much asymmetry is too much?}

\author[R. Livneh et al.]{
Ran Livneh,$^{1}$\thanks{E-mail: ran.livneh@weizmann.ac.il}
Boaz Katz$^{1}$
\\
$^{1}$Department of Particle Physics \& Astrophysics, The Weizmann Institute of Science, Rehovot 76100, Israel \\
}

\date{Accepted XXX. Received YYY; in original form ZZZ}

\pubyear{2021}

\begin{document}
\label{firstpage}
\pagerange{\pageref{firstpage}--\pageref{lastpage}}
\maketitle


\begin{abstract}
	
In a previous paper, we showed that the asymmetric ejecta produced by (zero impact parameter) head-on collisions of carbon-oxygen white dwarfs allow these progenitor models for Type Ia supernovae (SNe Ia) to cover the observed two-dimensional (2D) distribution of \ion{Si}{II} line depths (Branch plot). 
In this paper, we study the polarization signature associated with the 2D asymmetric ejecta of the collision model and a double-detonation model using similar \textsc{Tardis} radiative transfer simulations along different lines of sight with a spherical photosphere, combined with a new 3D Monte Carlo polarization code. 
We show that the polarization $Q$ can be parametrized as a product $Q=Q_{\max}Q_{\rm{x}}$ of a radial structure component $Q_{\max}$ which is insensitive to the model specifics and is shown to be universally around \mbox{$Q_{\max}\sim 5\%$}, and a cancellation component $Q_{\rm{x}}$ which depends on the asymmetry details. 
The continuum polarization is found to be low for both the collision and double-detonation models with $Q\sim 0.5\%$.
However, the irregular Si distribution in the 2D head-on collision model results in \ion{Si}{II} line polarization reaching $Q\sim 3\%$ ($Q_{\rm{x}} \lesssim 50\%$) in tension with observations (mostly $\lesssim 1.2\%$). 
In contrast, we show that the double-detonation model also covers the Branch plot, and yet results in low line polarization $Q\lesssim 0.7\%$ ($Q_{\rm{x}} \sim 10\%$) consistent with previous results and most SNe Ia.
These results strengthen the case for asymmetric explosions as progenitors of SNe Ia, emphasizing an additional requirement for large polarization cancellations to account for the low observed line polarizations. 

\end{abstract}

\begin{keywords}
supernovae: general -- radiative transfer -- polarization
\end{keywords}


\section{Introduction}

There is strong evidence that Type Ia supernovae (SNe Ia) are the product of thermonuclear explosions of white dwarfs (WDs), yet the nature of the progenitor systems and the mechanism that triggers the explosion remain long-standing open questions (see e.g. \citealt{Maoz2014,LivioMazzali2018,Soker2019}). 

The geometry of a SN explosion can help discriminate between various proposed progenitor systems. This geometry cannot be directly observed but leaves traces in the polarization signature that can be sensed and analyzed. Thus, modeling the continuum and line polarization of SN spectra can be an important tool in the quest for understanding SNe Ia progenitors.

Comparatively little observational polarization data exists, as spectropolarimetric observations consume significantly more resources than regular spectroscopic observations. Within the existing data, observed spectra show low continuum polarization of the order of $0.3\%$ and as high as $0.61\%$ \citep{Cikota2019}. These low polarization values have led to estimated ejecta asymmetries of $10\%-20\%$ \citep[e.g.][]{Hoflich1991, Kasen2003}. More significant polarization of up to $2\%$ is observed in strong spectral lines such as Silicon, Calcium, Sulfur, and Magnesium \citep{WangWheeler2008}, suggesting that asymmetries exist in the element distribution in the ejecta. Deviations from linear relations in the Stokes $Q-U$ plane as a function of wavelength have also been observed \citep{Cikota2019}, evidence that at least in some cases the SN ejecta does not exhibit axis-symmetry. Additionally, some correlations of \ion{Si}{II} line polarization with $\Delta m_{15}$ and with the pEW and velocity of the \ion{Si}{II} features have been reported \citep{Wang2007, Meng2017, Cikota2019}. 

Several studies regarding polarization in SNe Ia have been published. These range from the analysis of simple synthetic models of high-velocity feature polarization \citep{Kasen2003} to the addition of polarization modeling to radiative transfer simulators such as \textsc{Artis} \citep{Bulla2015}.  
In \citet{Bulla2016.2} a sub-Chandrasekhar mass double-detonation model \citep{Fink2010} and a Chandrasekhar mass delayed-detonation model \citep{Seitenzahl2013} were simulated and shown to produce low levels of \ion{Si}{II} line polarization ($\lesssim 1\%$) consistent with most of the observed SNe Ia ($\lesssim 1.2\%$), but inconsistent with higher polarization observations such as SN~2004dt ($\sim1.6\%$). In \citet{Bulla2016} the violent merger of a 1.1 and 0.9 $\textup{M}_\odot$ WD binary system \mbox{\citep{Pakmor2012}} was simulated, resulting in \ion{Si}{II} line polarizations of $\sim0.5\%-3.2\%$ over 35 viewing angles. This was concluded to be too high to explain the low polarization levels commonly observed in normal Type Ia supernovae. 

This last result may seem to imply that SNe Ia progenitor models displaying significant asymmetry should not be considered as viable. However, there are many indications that asymmetry in the ejecta may help explain various characteristics of SNe Ia, such as the variation in \ion{Si}{II} line velocity gradients, bi-modal and shifted nebular spectral lines and the presence of high velocity features \citep[e.g.][]{Maeda2010,Blondin2011,Childress2014,Kushnir2015,Dong2018,Maguire2018}.  
Specifically, \citet{Livneh2020} showed that the WD head-on collision model (e.g. \citealt{Rosswog2009}, \citealt{Raskin2010}, \citealt{Kushnir2013}) manages to reproduce the observed distribution of \ion{Si}{II} $\lambda 6355 \angstrom$ and $\lambda 5972 \angstrom$ line widths (Branch plot) due to the significant asymmetry of its ejecta. But, due to this asymmetry, it has been claimed that this model should exhibit a high degree of polarization inconsistent with observations \mbox{\citep[e.g.][]{Meng2017, LivioMazzali2018}}. 

In this paper, we study the polarization signatures produced by the 2D white dwarf head-on collision model using a new Monte Carlo polarization code. 
We find that the \ion{Si}{II} line polarizations reach $\sim 3\%$, in tension with observations. In contrast, we obtain low polarization values for an artificially extended double-detonation model from \citet{Townsley2019}, consistent with previous polarization calculations for double-detonation models and with most observations. 
We introduce a new method of analyzing the polarization as a product of a radial structure component and a cancellation component, giving further insight into our results, while at the same time elucidating the simple reason why SNe Ia typically present such low polarization signatures. 
Finally, we show that the double-detonation models can also reproduce SNe Ia characteristics associated with asymmetry, namely the Branch plot distribution, indicating there is no inherent contradiction between asymmetry and low polarization signatures. However, this asymmetry must be accompanied by a large degree of polarization cancellation to account for the low observed line polarization signatures.


\section{Methods}
\label{sec:Methods}


\subsection{Explosion Models}
\label{sec:Model}

\begin{figure*}
	\centering
	\includegraphics[clip, trim=0.0cm 1.0cm 0.0cm 0.7cm, width=\textwidth]{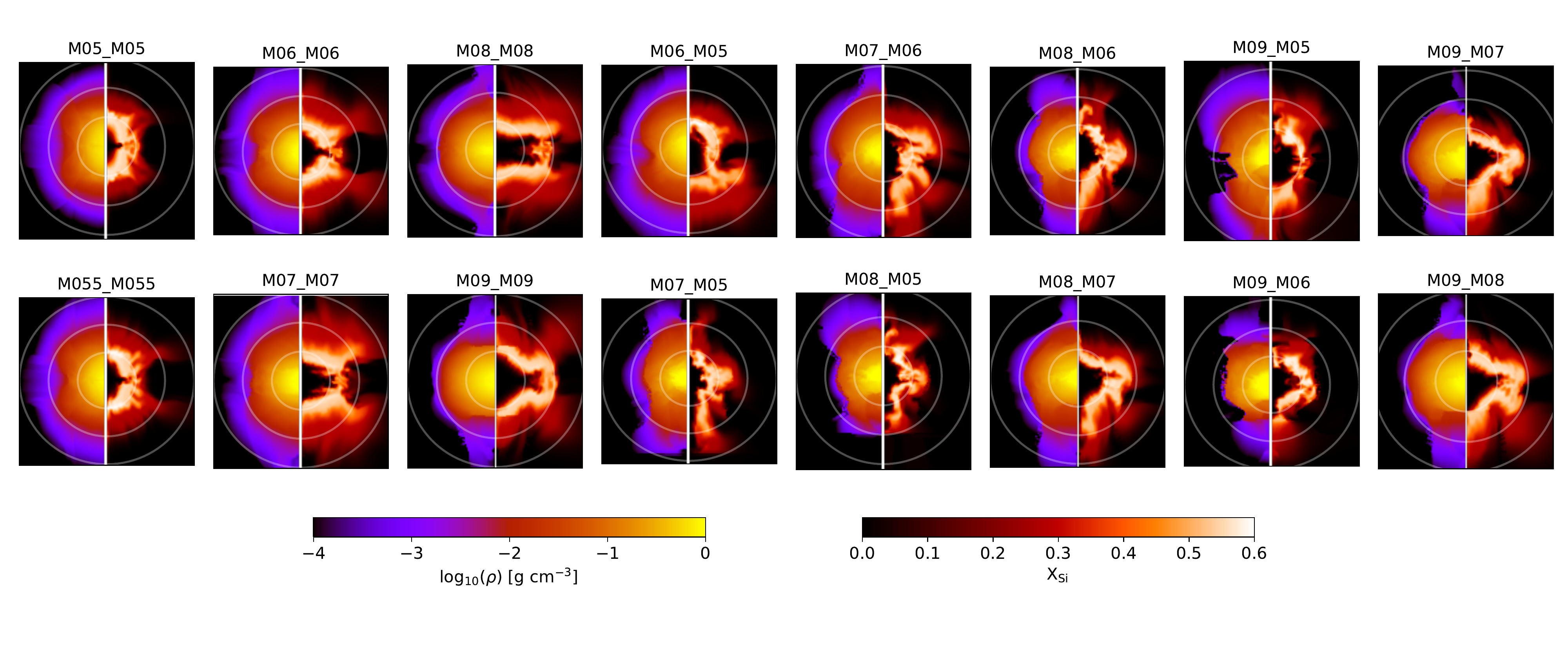}
	
	\caption{Ejecta of 2D head-on collision models at 100 sec after detonation. Collision takes place on the symmetry axis. The left part of each image shows the density, while the right side shows the Si abundance. Concentric circles are shown at $1$, $2$ and $3\times10^4$~km/s. }
		
	\centering
	\label{fig:ColModels}
\end{figure*}

Two-dimensional head-on (zero impact parameter) WD collision ejecta were taken from \citet{Kushnir2013}, where equal and non-equal mass CO-WDs with masses 0.5, 0.6, 0.7, 0.8, 0.9 and 1.0 $\textup{M}_\odot$ were simulated, resulting in explosions that synthesize $^{56}$Ni masses in the range of $0.1\,\textup{M}_\odot$ to $1.0\,\textup{M}_\odot$. Fig.~\ref{fig:ColModels} shows the density and Si abundance of all collision models used in this study, summarized in Table~\ref{table:collisions}. The luminosity $L_{\rm bol}^{\max}$ and rise time $t_{\rm rise}($bol$)$ were taken from global 2D LTE radiation transfer simulations of the same ejecta performed by Wygoda, N. (private communication) using a 2D version of the \textsc{Urilight} radiation transfer code \citep[][Appendix A]{Nahliel2}.


\begin{table}
	\centering
	\begin{tabular}{ |c|c|c|c|c|c| } 
		\hline
		Model & $M_1$              & $M_2$             & $M(^{56}$Ni$)$    & $t_{\rm rise}($bol$)$ & $L_{\rm bol}^{\max}$ \\
		& ($\textup{M}_\odot$)  & ($\textup{M}_\odot$) & ($\textup{M}_\odot$)  & (days) & (erg/s) \\
		\hline
		\multicolumn{6}{|c|}{Head-on collision models} \\
		\hline
		M05\_M05 & 0.50 & 0.50 & 0.10 & 14.8 & 2.98 (42)\\
		M055\_M055 & 0.55 & 0.55 & 0.22 & 15.7 & 5.56 (42)\\
		M06\_M05 & 0.60 & 0.50 & 0.27 & 15.3 & 6.54 (42)\\
		M06\_M06 & 0.60 & 0.60 & 0.33 & 16.0 & 7.64 (42)\\
		M07\_M05 & 0.70 & 0.50 & 0.26 & 15.7 & 6.51 (42)\\
		M07\_M06 & 0.70 & 0.60 & 0.38 & 16.0 & 8.81 (42)\\
		M07\_M07 & 0.70 & 0.70 & 0.56 & 15.9 & 1.25 (43)\\
		M08\_M05 & 0.80 & 0.50 & 0.29 & 16.2 & 7.32 (42)\\
		M08\_M06 & 0.80 & 0.60 & 0.38 & 16.3 & 9.54 (42)\\
		M08\_M07 & 0.80 & 0.70 & 0.48 & 16.5 & 1.17 (43)\\
		M08\_M08 & 0.80 & 0.80 & 0.74 & 15.5 & 1.67 (43)\\
		M09\_M05 & 0.90 & 0.50 & 0.69 & 15.6 & 1.34 (43)\\
		M09\_M06 & 0.90 & 0.60 & 0.50 & 16.5 & 1.26 (43)\\
		M09\_M07 & 0.90 & 0.70 & 0.51 & 16.7 & 1.23 (43)\\
		M09\_M08 & 0.90 & 0.80 & 0.54 & 17.1 & 1.27 (43)\\
		M09\_M09 & 0.90 & 0.90 & 0.78 & 16.8 & 1.74 (43)\\
		\hline
	\end{tabular}
	\caption{Properties of SNe Ia 2D head-on collision models adapted from \citet{Kushnir2013}. Numbers in parentheses correspond to powers of ten. $L_{\rm bol}^{\max}$ and $t_{\rm rise}($bol$)$ represent averages over all viewing angles.}
	\label{table:collisions}
\end{table}



\begin{figure*}
	\centering
	
	\includegraphics[clip, width=\textwidth, trim=0.0cm 1.0cm 0.0cm 0.7cm] {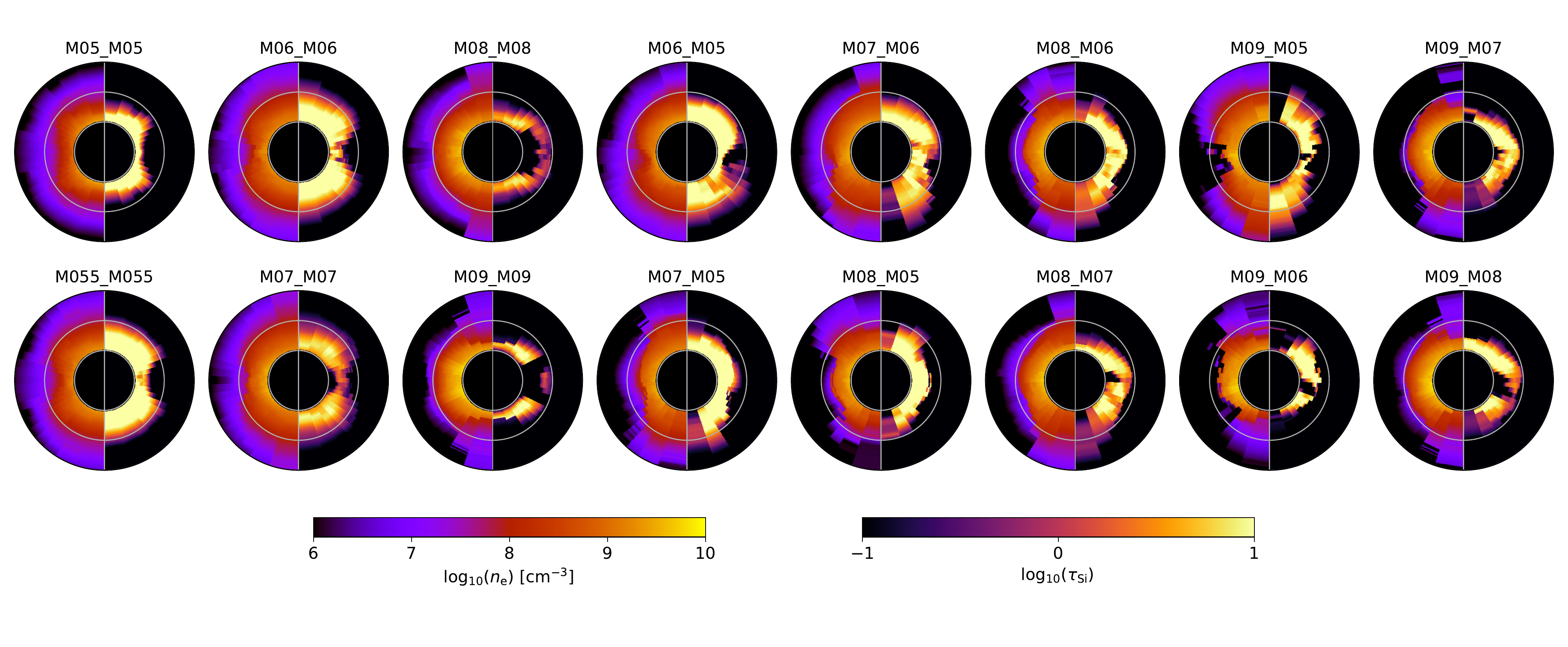}
	
	\caption{Results of \textsc{Tardis} runs on ejecta of 2D head-on collision models at 16d after detonation. Each viewing angle is run independently and then combined into the shown 2D map. Collision takes place on the symmetry axis. The left part of each image shows the electron number density, while the right side shows the optical depth for the \ion{Si}{II} $\lambda 6355 \angstrom$ line (values $\tau_{\rm{Si}}>10$ are saturated). Concentric circles are shown at $1$, $2$ and $3\times10^4$~km/s.}
	
	\centering
	\label{fig:ColModels_tau}
\end{figure*}


In addition to the 2D head-on collision models, we study a 2D double-detonation model from \citet{Townsley2019}. The left panel of Fig.~\ref{fig:Townsley_ejecta} shows the density and Si abundance of this model. We extend the published model synthetically by running the radiative transfer simulation on its ejecta with varying target luminosities in the range $3\times10^{42}$ to $1.4\times10^{43}$ erg/s. Note that this extension is artificial and should only serve as an initial proof of concept, while a more complete study of hydro-dynamically simulated double-detonation models \citep[e.g.][]{Boos2021,Shen2021} is beyond the scope of this paper. 


\subsection{Ionization and excitation}
\label{sec:Ionization_and_excitation}

Spectral polarization calculations are carried out in two main steps. First, we use a radiative transfer simulation to calculate the electron densities and the optical depths of the various lines in the ejecta. Next, we use these as input to a new polarization simulator that propagates photon packets while taking into account the changes in their polarization as they are scattered from electrons and excited ions. 

For the first step we use the photospheric spectral synthesis code \textsc{Tardis} \citep{Tardis}. \textsc{Tardis} is a one-dimensional (1D) simulator, and to use it we divide the two-dimensional (2D) ejecta into 21 viewing angles, each simulated separately. A division into 42 viewing angles was tested but was found not to have an appreciable effect on results. 
For each viewing angle, a 1D model was generated using the density and abundance values sampled along the section. 
The degree of line polarization varies with the observation epoch, but for most observed SNe, at the time of B-band maximum it is close to the peak polarization \citep[e.g.][]{Bulla2016, Cikota2019}. For the purpose of this study, we simulate the model ejecta at 16d (14d) after the explosion for the collision model (double-detonation model). 
A photospheric velocity of $10,000$~km/s is assumed for all models and viewing angles unless indicated otherwise. The sensitivity to this choice is explored in Section~\ref{sec:Results}. For the \textsc{Tardis} simulations, we use a radial resolution of 500~km/s for all models, close to the output resolution of the original hydrodynamical simulations. In our study, we use \textsc{Tardis} with the most detailed \texttt{macroatom} model and the \texttt{dilute-lte} excitation mode. As an exception, the excitation levels of \ion{Ca}{II}, \ion{S}{II}, \ion{Mg}{II} and \ion{Si}{II} ions are calculated with a "full NLTE" treatment. As shown in \citet{Tardis}, this has a significant effect on the pEW of the \ion{Si}{II} $5750\angstrom$ feature. Further details are provided in \citet{Livneh2020}. 

Next, we combine the 21 viewing angles back into a 2D map of the electron densities and line optical depths (see Fig.~\ref{fig:ColModels_tau}). This map is used as a basis for the polarization simulator (described below), along with the average of the photospheric temperatures $\overline{T_{\rm{ph}}}$ over all viewing angles. 


\subsection{Polarization}
\label{sec:Polarization}

To enable an in-depth investigation of polarization aspects, a new 3D Monte Carlo polarization simulator was developed. Unpolarized photon packets with wavelengths sampled from a blackbody distribution with $T=\overline{T_{\rm{ph}}}$ are launched from random points on the photosphere in random directions. Emission points on the photosphere are chosen isotropically, while the propagation direction is chosen using $\mu=\sqrt{z},z\in \left(0,1\right]$, as appropriate for zero limb darkening. The packets are then propagated through a given 3D ejecta, undergoing scattering either by electrons or by excited ions in regions of Sobolev resonance \citep{Sobolev1960}. 

To find when continuum and interaction events occur, we use a method outlined by \citet{MazzaliLucy1993}. For each photon packet, a maximum optical depth $\tau_{\max}$ is randomly chosen in accordance with the $e^{-\tau}$ attenuation law: $\tau_{\max} = -\ln(z), z \in (0,1]$. The packet proceeds while accumulating opacity due to the electron density $\tau_{\rm{es}}=\sigma_{\rm{T}}n_{\rm{e}}(1-\mu v/c)s$ along a path length $s$ \citep{Lucy2005}. This process continues until one of the following occurs: (a) The packet reaches the inner boundary (which is identical to the $v_{\rm{ph}}$ used in \textsc{Tardis}) and is absorbed. (b) The packet reaches the outer boundary and is emitted. (c) The accumulated $\tau$ due to opacity is greater than  $\tau_{\max}$, indicating an electron scattering event. (d) The packet reaches resonance with an absorption line. In this case the optical depth of the line (previously computed using \textsc{Tardis}) is added to $\tau$. If the new $\tau>\tau_{\max}$, a line scattering event occurs and a new $\tau_{\max}$ is generated. 


\begin{figure}
	\centering
	
	\includegraphics[clip, trim=0.0cm 1.0cm 0cm 0cm, width=0.9\columnwidth] {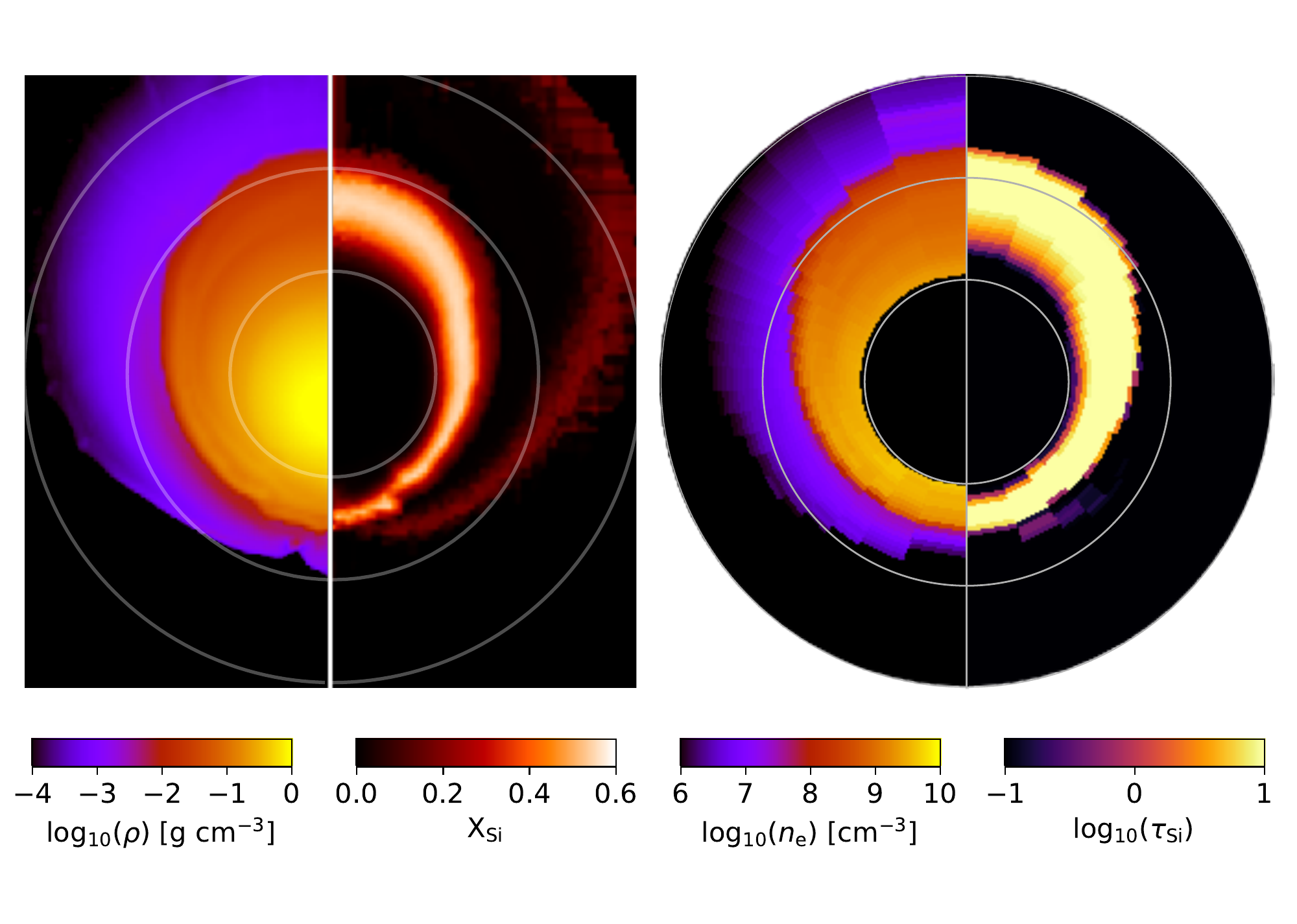}
	
	\caption{Left: Same as Fig.~\ref{fig:ColModels} for a double-detonation model from \citet{Townsley2019}. Right: Same as Fig.~\ref{fig:ColModels_tau} for \textsc{Tardis} runs on ejecta of double-detonation model with $\log(L/\rm{L_{\odot}})=9.20$ at 14d after detonation.}
	
	\centering
	\label{fig:Townsley_ejecta}
\end{figure}


The polarization of a beam of radiation is characterized by the four-dimensional Stokes vector $S = (I, Q, U, V)$. The $I$ component measures the total intensity, while $Q$ and $U$ measure the degree of linear polarization in 2 directions separated by $45^{\circ}$. $V$ measures circular polarization which has never been observed in SNe Ia and is thus neglected \citep{Chandrasekhar1960}. 

Electron scattering is treated as outlined in \citet{CodeWhitney1995,Whitney2011}. The location of the scattering event is identified with a velocity resolution of 10~km/s. The interaction changes the polarization of the photon packet, transforming its Stokes vector according to a scattering matrix. The probability of scattering into various angles is determined by sampling a probability distribution dependent on the Stokes vector and the scattering angle. We neglect the effects of relativistic aberration.

In contrast, resonance line scattering depolarizes the photon packet and re-emits it with the same co-moving frame frequency but a new direction of propagation determined by randomly sampling an isotropic distribution $\mu=-1+2z,z\in \left[0,1\right]$. Only the simplest bound-bound scattering is assumed in the MC polarization simulator, disregarding bound-free and free-free interactions and neglecting the possibility of internal state transitions within the scattering ions.

The photon packets are tracked until they exit the ejecta and are subsequently collected in bins forming both a flux and a polarization spectrum. Flux is collected in $5\angstrom$ bins, while the polarization is collected in $25\angstrom$ bins. This step assumes a 2D axis-symmetric configuration. We focus on the spectral region of the \ion{Si}{II} features, collecting photon packets within the wavelength range $5300\angstrom-7100\angstrom$. Since the polarization signature is weak, an order of magnitude of $10^9$ packets are used per model.  
After extracting the polarization spectrum, the $U$~polarization is verified to be identically zero (up to numerical error) as expected from axis-symmetric ejecta where the projection of the symmetry axis on the observed plane is along the y-axis.
 
For the \ion{Si}{II} polarization runs, only lines with $\tau>0.1$ within the wavelength range $5600\angstrom-6600\angstrom$ are included. Continuum polarization values are extracted from the wavelength range between $5400\angstrom$ and $5450\angstrom$, where the scattering lines have no effect.
Line polarization is computed as the maximal absolute polarization obtained in the range $5950\angstrom-6200\angstrom$ after the continuum polarization is subtracted. Subtraction of the continuum polarization simulates the method used in \citet{Cikota2019}, in which wavelet decomposition is used to remove the continuum component - whether it originates from external inter-stellar polarization (ISP) or intrinsic continuum polarization.
Verification of this polarization simulator is provided in Appendix \ref{sec:SimulatorVerification}.

We emphasize that our modeling assumes a spherical photosphere, taking into account only the effects of the electron and ion distributions, not the spatial configuration of the radioactive sources. In reality, the emitting surface can have a convoluted shape depending on the configuration of IGEs in the ejecta. This surface can also vary with wavelength due to variations in thermalization depth. For example, extreme ellipsoidal toy models have been shown to exhibit sign reversals from shorter to longer wavelengths due to this effect \citep[e.g.][]{Bulla2015}.  
Additionally, due to the limitations of the 1D radiative transfer simulation, the photospheric velocity $v_{\rm{ph}}$ must be input by hand into the model. Since the value of $v_{\rm{ph}}$ determines the optical depth for polarizing collisions with electrons, it has an appreciable effect on the degree of polarization, explored in Section~\ref{sec:Results}. 


\section{Results}
\label{sec:Results}

\begin{figure}
	\centering
	\includegraphics[clip, trim=1.0cm 1.0cm 1.6cm 1.0cm, width=\columnwidth]{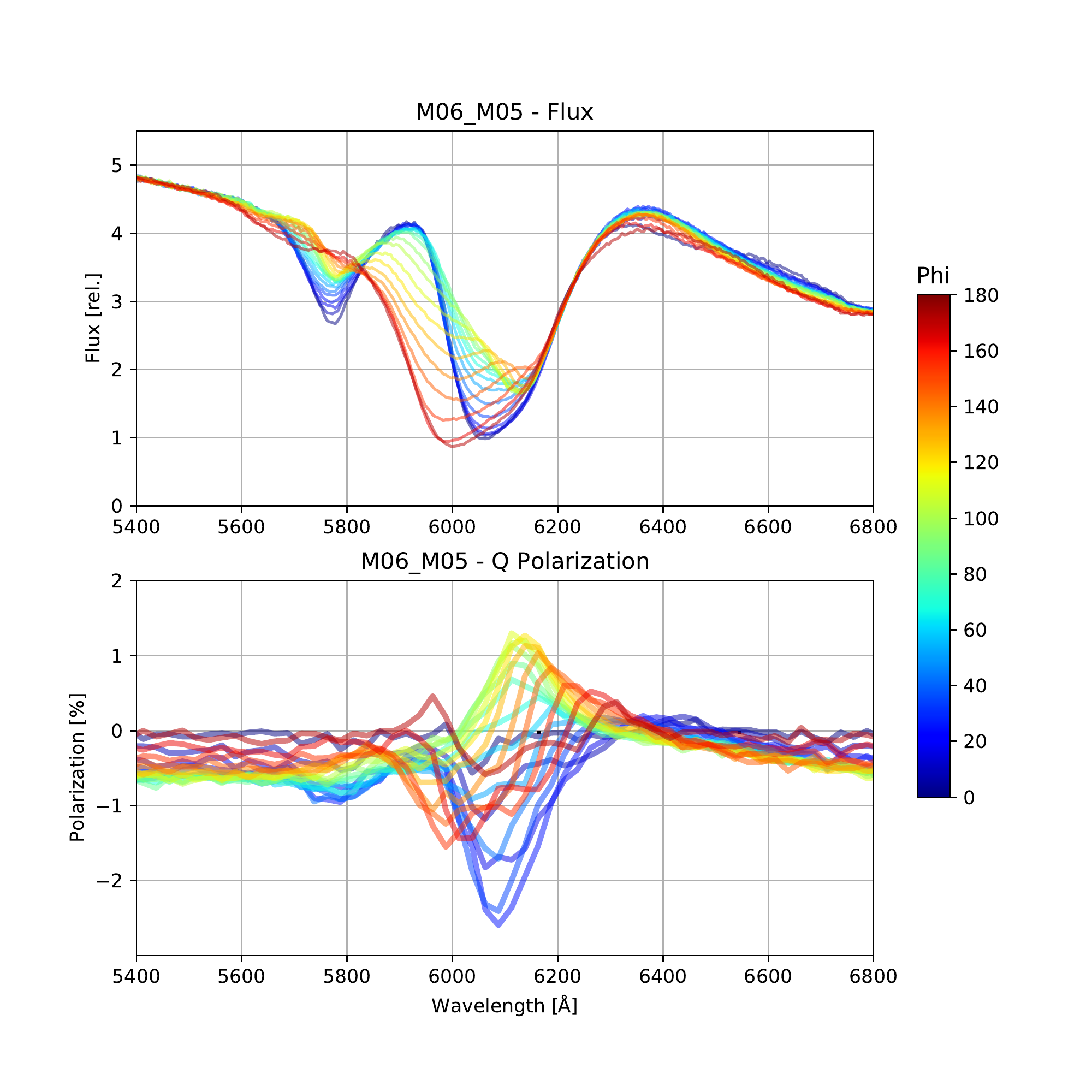}
	
	\caption{Flux and $Q$ polarization of an example 2D head-on collision model. Each color represents a different viewing angle $\phi$ with respect to the z-axis. $U$~polarization is identically zero due to the axis-symmetry of the model. Only lines with $\tau>0.1$ within the wavelength range $5600\angstrom-6600\angstrom$ are included. Flux is collected in $5\angstrom$ bins, while the polarization is collected in $25\angstrom$ bins. Continuum polarization values can be extracted from the wavelength range between $5400\angstrom$ and $5450\angstrom$, where the scattering lines have no effect.}
	\centering
	\label{fig:spectrum_example}
\end{figure}
\begin{figure*}
	\centering
	\includegraphics[clip, width=\textwidth, trim=1.0cm 1.0cm 1.0cm 0.0cm] {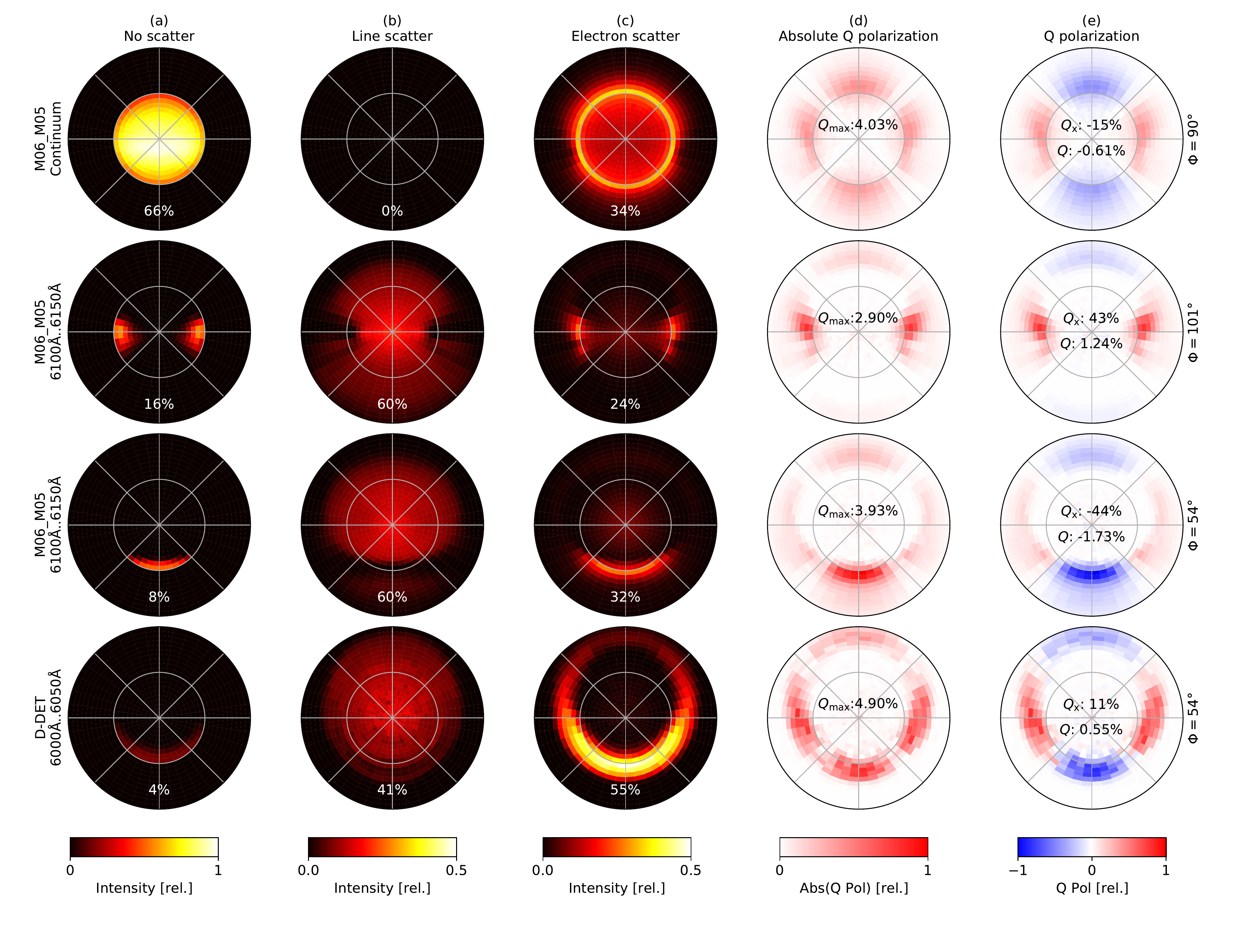}
	
	\caption{Example of formation of continuum and line polarization in one 2D head-on collision model (\texttt{M06\_M05}) and one double-detonation model ($\log(L/\rm{L_{\odot}})=9.20$). Viewing angles $\phi$ are shown on the right. The images show the projection on a plane of photon packets within a selected wavelength range (shown on left) reaching the observer. The top row is in a continuum spectral region with no lines. The other rows are in the spectral region of the peak absorption of the \ion{Si}{II} $\lambda 6355 \angstrom$ line. The columns show: (a)-(c) Intensity map and percentage of the flux arriving to the observer from different locations on the projected disk according to their last interaction type. (d)-(e) The absolute and signed $Q$ polarization of light arriving from different locations on the projected disk. The resulting total $Q$ polarization and its components $Q_{\max}$ and $Q_{\rm{x}}$ are also displayed (see Section~\ref{sec:Analysis}). The radial range of all panels is $0-20,000$~km/s. }
	
	\centering
	\label{fig:scatters}
\end{figure*}

A detailed example of the resulting polarization for one head-on collision model (\texttt{M06\_M05}) is shown in Figs.~\ref{fig:spectrum_example},~\ref{fig:scatters}. The results of the maximum \ion{Si}{II} line polarization for all collision models are shown in Figs.~\ref{fig:max_vs_ni56},~\ref{fig:max_vs_ni56_vph}, while the continuum polarization is shown in Fig.~\ref{fig:cont_vs_ni56_vph}. Polarization results for the double-detonation model are shown in Figs.~\ref{fig:scatters},~\ref{fig:max_vs_ni56_vph} and \ref{fig:cont_vs_ni56_vph}.


The degree and angle of polarization of emitted light are determined primarily by the location and type of its last interaction with the ejecta. Fig.~\ref{fig:scatters} shows an example of continuum and line polarization formation in one 2D head-on collision model (\texttt{M06\_M05}) and one double-detonation model. Columns (a)-(c) show the intensity of photon packets arriving to the observer from different locations on the projected disk, according to their last interaction type. Column (e) shows the positive and negative contributions of this flux to the overall $Q$ polarization. Note that we analyze $Q$ for our axis-symmetric model without loss of generality, since the analysis is carried out independently for each viewing angle and wavelength, and thus even in a non-axis-symmetric case, the orientation can be chosen such that $U=0$. 

The first row in this figure shows the formation of continuum polarization for a viewing angle of $\phi= 90^{\circ}$. The electron cloud close to the photosphere is slightly denser closer to the symmetry axis (top and bottom) compared to the sides (see also Fig.~\ref{fig:ColModels_tau}). As a result, the negative contributions to the polarization are slightly larger than the positive contributions, resulting in an overall negative continuum polarization of $Q=-0.61\%$. 

In the second and third row, two examples of the formation of line polarization in the wavelength range $6100\angstrom-6150\angstrom$ are shown, both of the same collision model but from 2 viewing angles. The second row shows a viewing angle of $\phi= 101^{\circ}$. In this case, the \ion{Si}{II} ions obstruct and depolarize the emitted polarized light mostly at the top and bottom portions of the projected plane. This allows mostly polarized light from the sides to pass, resulting in positive polarization of $1.24\%$. The third row shows a viewing angle of $\phi= 54^{\circ}$. In this case, the \ion{Si}{II} ions obstruct the emitted polarized light almost everywhere except on the bottom portion of the projected plane, resulting in negative polarization of -$1.73\%$. The bottom row shows an example of a double-detonation model ($\log(L/\rm{L_{\odot}})=9.20$, $\phi= 54^{\circ}$): here the obstructing \ion{Si}{II} ions allow polarized light to still arrive from all quadrants, resulting in a large degree of cancellation leading to a low polarization of $Q=0.55\%$.


\subsection{Line polarization}
\label{sec:ResultsLinePolarization}

\begin{figure}
	\centering
	\includegraphics[clip, width=\columnwidth]{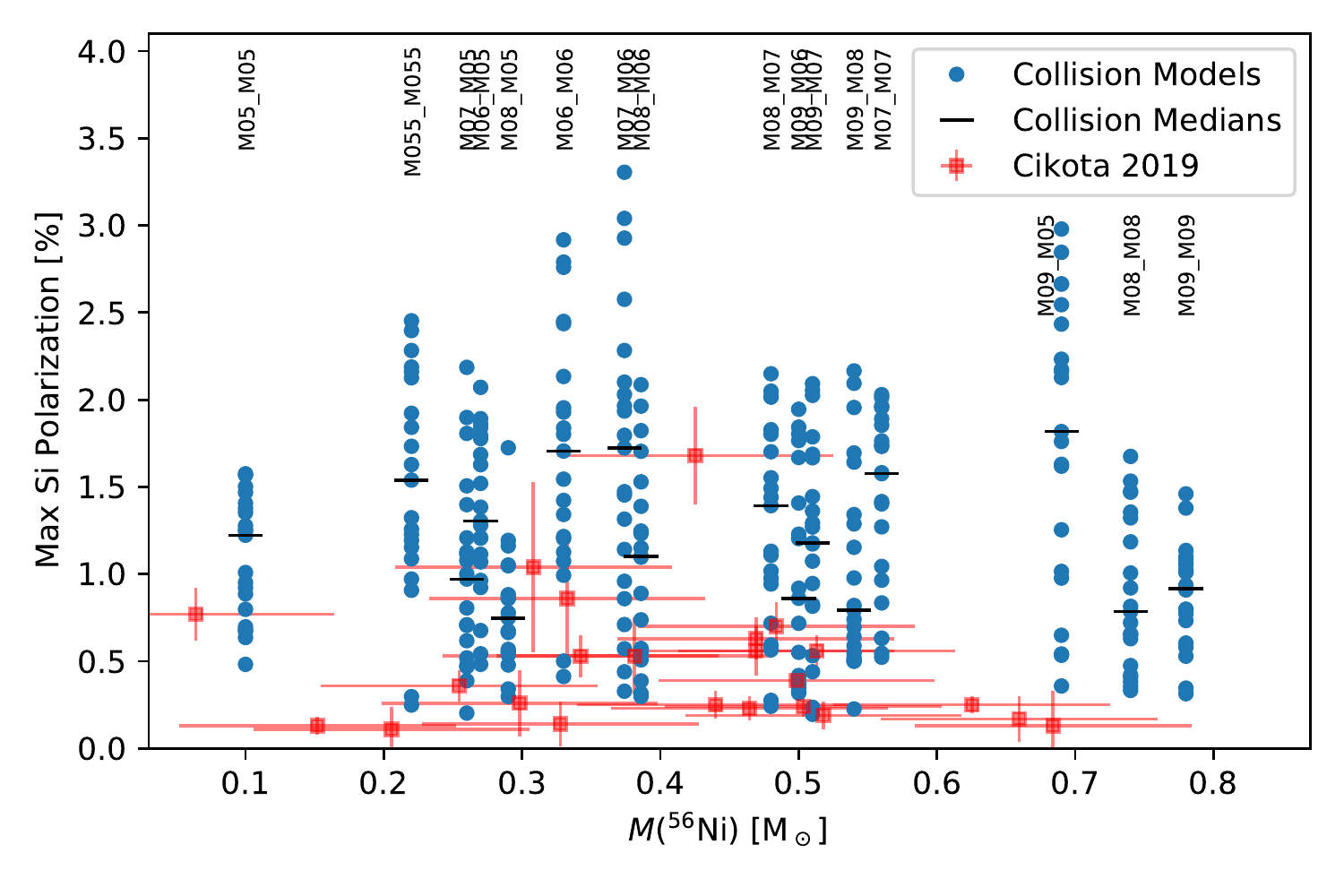}
	
	\caption{Maximum \ion{Si}{II} line polarization for the 2D head-on collision models with $v_{\rm{ph}}=10,000$~km/s at 16d (blue circles). The median Si polarization for each model is indicated by a black line. Red squares indicate \ion{Si}{II} line polarizations from \citet{Cikota2019} -- data are taken for the epoch closest to but before 0d, at $25\angstrom$ binning. Error bars for $\rm{M(^{56}Ni})$ are shown for illustration and taken to be $\sim0.1\rm{M_{\astrosun}}$ due to conversion using an approximate linear relation $\rm{M(^{56}Ni})/M_{\astrosun}=1.016-0.488\Delta m_{15}$, derived from data in \citet{Nahliel2}.}
	\centering
	\label{fig:max_vs_ni56}
\end{figure}

The distribution of maximum \ion{Si}{II} line polarization values for 2D head-on collision models with $v_{\rm{ph}}=10,000$~km/s at all viewing angles as a function of synthesized $^{56}$Ni mass is shown in Fig.~\ref{fig:max_vs_ni56}, compared to observations from \citet{Cikota2019}, with $\Delta m_{15}$ values converted to $\rm{M(^{56}Ni})$ using an approximate linear relation $\rm{M(^{56}Ni})/M_{\astrosun}=1.016-0.488\Delta m_{15}$, derived from data in \citet{Nahliel2}. The resulting line polarizations tend to be larger than those observed by a factor of 2-3.


\begin{figure}
	\centering
	\includegraphics[clip, width=\columnwidth]{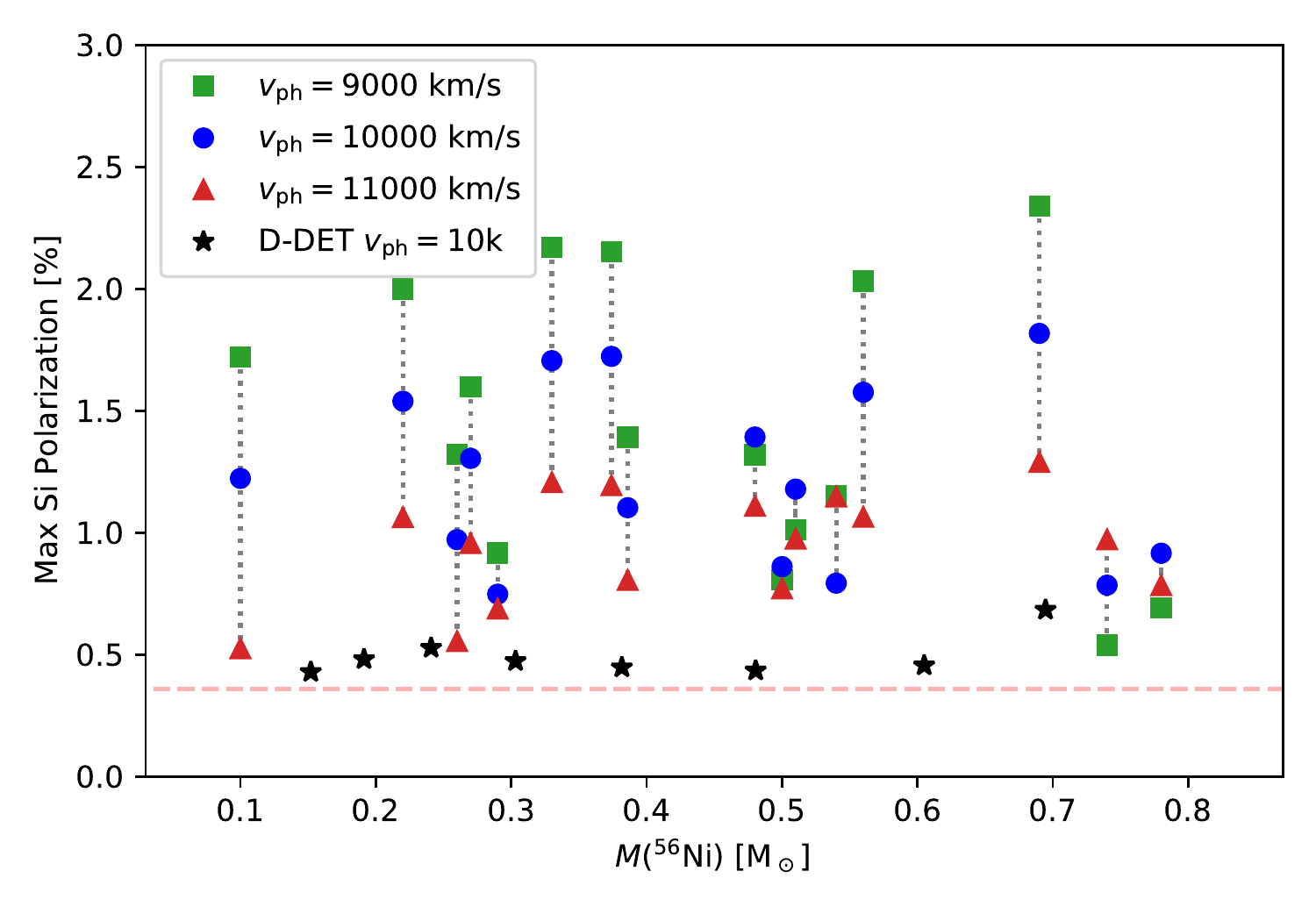}
	
	\caption{Maximum \ion{Si}{II} line polarization for the 2D head-on collision models: median line polarization values are shown for different choices of $v_{\rm{ph}}$. Also shown are the median line polarizations for the double-detonation models for $v_{\rm{ph}}=10,000$~km/s ($M(^{56}$Ni$)$ is computed from $L$ using Arnett's rule $L_{\max}=2.0\times10^{43}\times[M(^{56}$Ni$)/\textup{M}_\odot]$~erg/s). The scatter of the individual viewing angles for the double-detonation models can be seen in Fig.~\ref{fig:Breakdown}. The red dashed line indicates the median of line polarizations from \citet{Cikota2019} (as in Fig.~\ref{fig:max_vs_ni56}).}
	
	\centering
	\label{fig:max_vs_ni56_vph}
\end{figure}

Due to the limitations of the radiative transfer simulation (see Section~\ref{sec:Ionization_and_excitation}), the photospheric velocity $v_{\rm{ph}}$ must be input by hand to the model. Since changing $v_{\rm{ph}}$ affects the optical depth for polarizing collisions with electrons, it is crucial to check the effect of this parameter on the polarization results. Fig.~\ref{fig:max_vs_ni56_vph} shows the effect of varying $v_{\rm{ph}}$ between $9,000$ and $11,000$~km/s on the median of the resulting line polarization. Both the \textsc{Tardis} radiative transfer simulation and the polarization simulation were updated to use the new value of $v_{\rm{ph}}$. For most models there is a substantial effect. Reducing $v_{\rm{ph}}$ to $9,000$~km/s (which is the value used in \citet{Livneh2020}) further exacerbates the tension between the model and the observed values. Increasing $v_{\rm{ph}}$ to $11,000$~km/s could decrease the tension somewhat, but this choice would affect the velocity distribution of the \ion{Si}{II} lines in a way that would not be consistent with observations. 
Also shown in Fig.~\ref{fig:max_vs_ni56_vph} are the median maximum \ion{Si}{II} line polarizations obtained for the double-detonation models. The values are much lower and closer to the median of the observed values (dashed line), with negligible dependence on the luminosity. 


\subsection{Continuum polarization}
\label{sec:ResultsContinuumPolarization}


\begin{figure}
	\centering
	\includegraphics[clip,  width=\columnwidth]{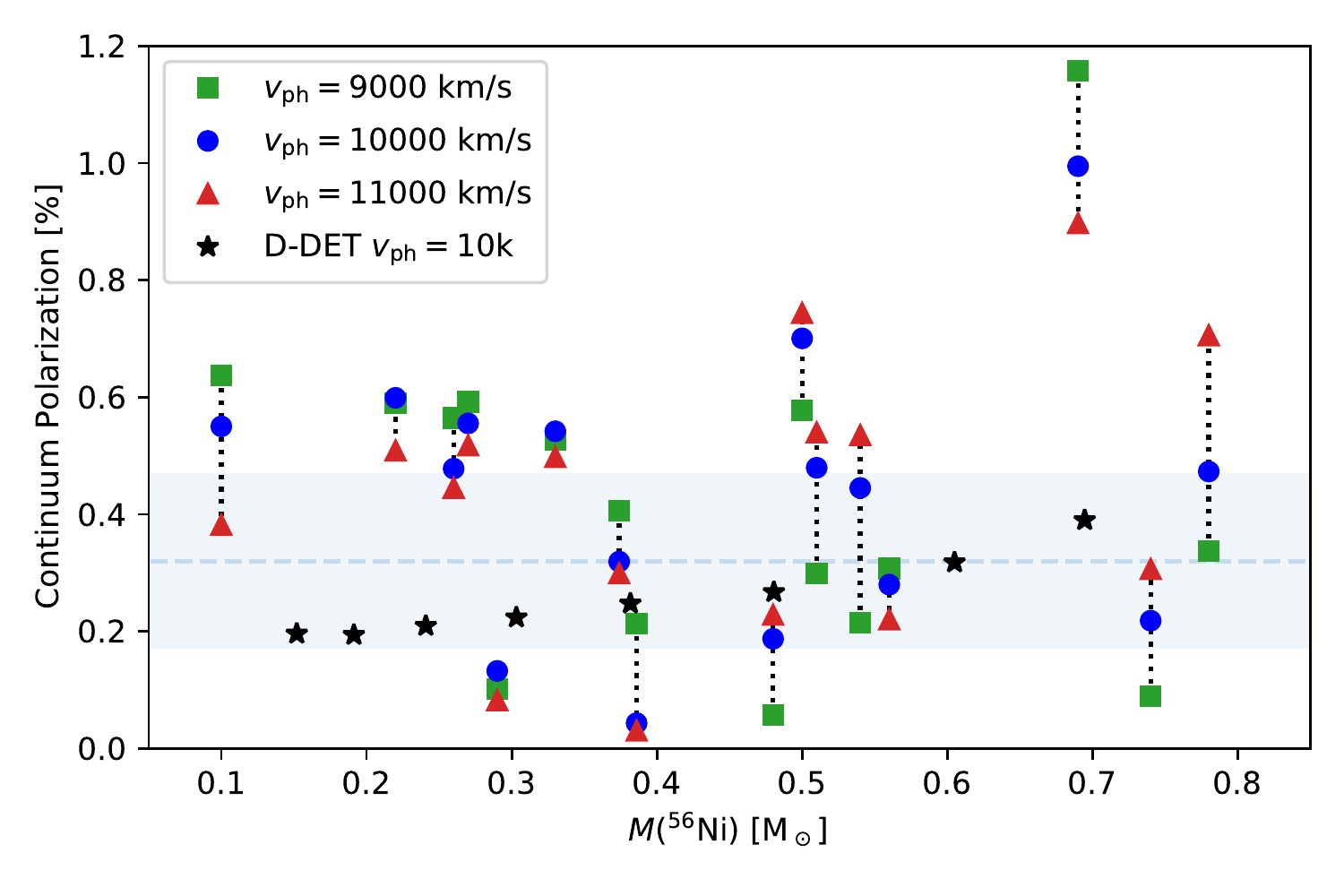}
	
	\caption{Continuum polarization for the 2D head-on collision models. Median continuum polarization values are shown for different choices of $v_{\rm{ph}}$. Also shown are the median continuum polarizations for the double-detonation models for $v_{\rm{ph}}=10,000$~km/s ($M(^{56}$Ni$)$ is computed from $L$ using Arnett's rule $L_{\max}=2.0\times10^{43}\times[M(^{56}$Ni$)/\textup{M}_\odot]$~erg/s). The dashed line and shaded area represent the observed continuum polarization of 6 SNe with low reddening ($0.32\pm0.15\%$) from \citet{Cikota2019}.}
	
	\centering
	\label{fig:cont_vs_ni56_vph}
\end{figure}


Intrinsic continuum polarization values are difficult to reliably extract from observations due to ISP and other sources of continuum polarization \citep[e.g.][]{WangWheeler2008}. Additionally, continuum polarization simulation results are expected to depend more strongly on the configuration of the radioactive sources, which we do not take into account in our simulations. Still, for completeness, we present the continuum polarization values obtained for the two models using our method. 

In the example given in Fig.~\ref{fig:spectrum_example}, we can see that the continuum polarization (at the edges of the shown wavelength range) tends to zero when observed along the line of the collision ($\phi=0,\pi$), and is maximal when observed perpendicular to the collision axis ($\phi=\pi/2$). 
The median of the resulting continuum polarization for each collision model as a function of synthesized $^{56}$Ni mass can be seen in Fig.~\ref{fig:cont_vs_ni56_vph} for $v_{\rm{ph}}$ between $9,000$ and $11,000$~km/s. Also shown are the median continuum polarization values of the double-detonation models. These values are compared to observations from \citet{Cikota2019}, where continuum polarization values for six low-reddening SNe were measured to be $0.32\pm0.15\%$ with an upper limit of $0.61\%$. Both models display low continuum polarizations of $\sim0.5\%$, consistent with observations. 

The relatively low continuum polarization values highlight the fact that the distribution of free electrons in the ejecta of the collision models is relatively symmetric when compared with the distribution of \ion{Si}{II} ions, as can also be seen in Fig.~\ref{fig:ColModels_tau}. It is also interesting to note that the direction of change of the polarization as a function of $v_{\rm{ph}}$ is not constant. This emphasizes the complex dependence of polarization signals on the structure of the ejecta. 


\section{Polarization Analysis}
\label{sec:Analysis}

As column (e) of Fig.~\ref{fig:scatters} demonstrates, the positive and negative contributions to the total $Q$ polarization generally arrive from different quadrants on the projected disk. The total $Q$ polarization can be written as $Q=\left(Q^{+}-Q^{-}\right)/I$, where $Q^{+}$($Q^{-}$) represents the absolute value of the positive (negative) contribution to the polarization and $I$ is the total flux. We expand this to:
\begin{equation}
	Q=\frac{Q^{+}+Q^{-}}{I}\frac{Q^{+}-Q^{-}}{Q^{+}+Q^{-}} \equiv Q_{\max}Q_{\rm{x}} 
	\label{eq:breakdown}
\end{equation}

\begin{figure}
	\centering
	\includegraphics[clip,  width=\columnwidth]{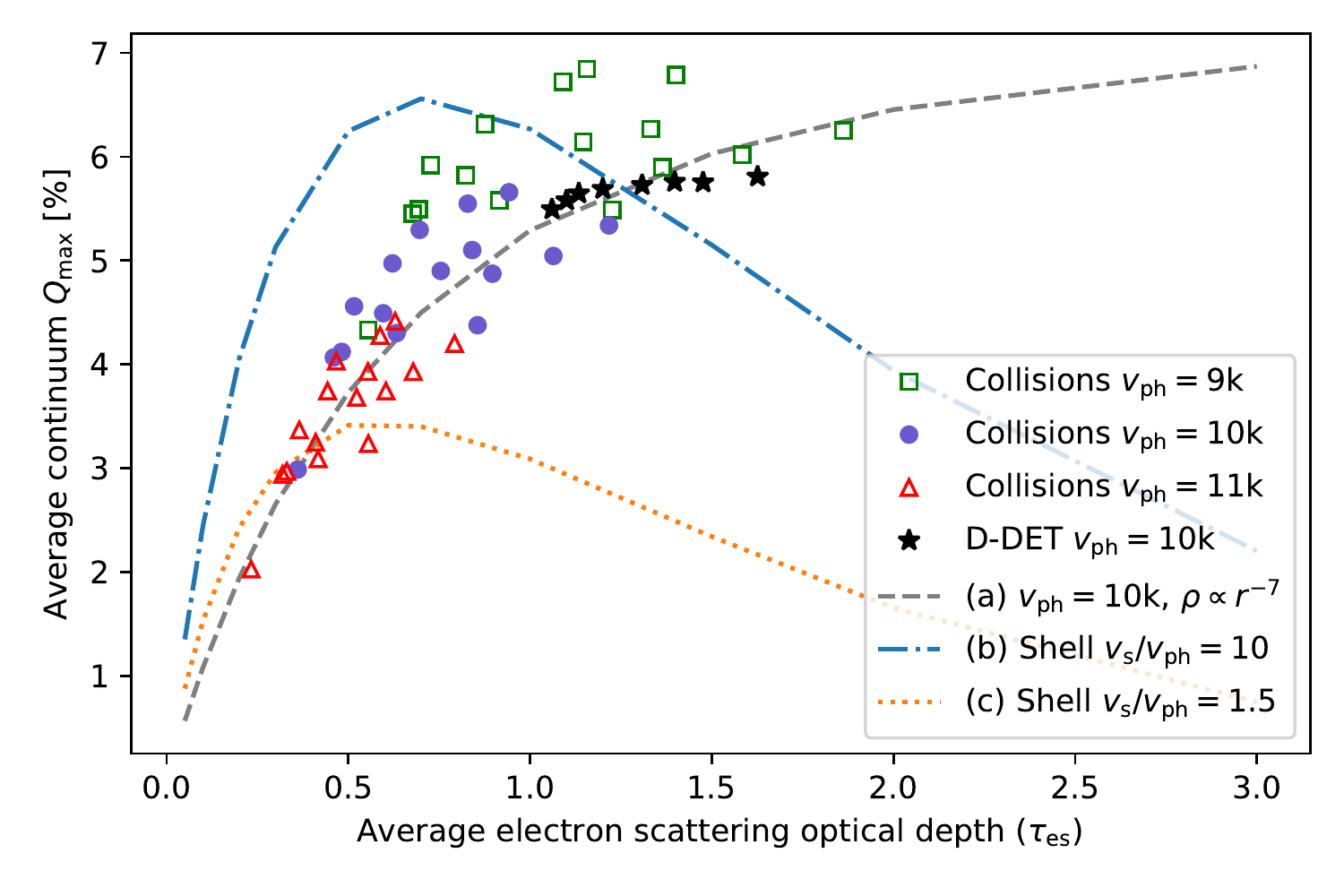}
	\caption{Average continuum $Q_{\max}$ values for various models as a function of average optical depth (in the radial direction) from the inner boundary surface to infinity due to electron scattering. Each point represents the average continuum polarization of one model over all viewing angles. Collision models are shown for 3 choices of $v_{\rm{ph}}$ and double-detonation models (see Section~\ref{sec:Model}) are shown with $v_{\rm{ph}}=10,000$~km/s. Overlaid are synthetic models: (a) $v_{\rm{ph}}=10,000$~km/s with a power-law electron scattering atmosphere $\rho\propto r^{-7}$ ; (b) An approximation of a point source: $v_{\rm{ph}}=1,000$~km/s with a $100$~km/s shell at $v_{\rm{s}}=10,000$~km/s; and (c) $v_{\rm{ph}}=10,000$~km/s with a $100$~km/s shell at $v_{\rm{s}}=15,000$~km/s.}
	
	\centering
	\label{fig:Qmax}
\end{figure}


This breaks the total polarization into 2 components: $Q_{\max}=\frac{Q^{+}+Q^{-}}{I}$ represents the total $Q$ polarization assuming contributions from all locations were summed with an absolute value (see column (d) of Fig.~\ref{fig:scatters}), while $Q_{\rm{x}}=\frac{Q^{+}-Q^{-}}{Q^{+}+Q^{-}}$ represents the cancellation of the positive and negative components.

\begin{figure}
	\centering
	\includegraphics[clip,  width=\columnwidth]{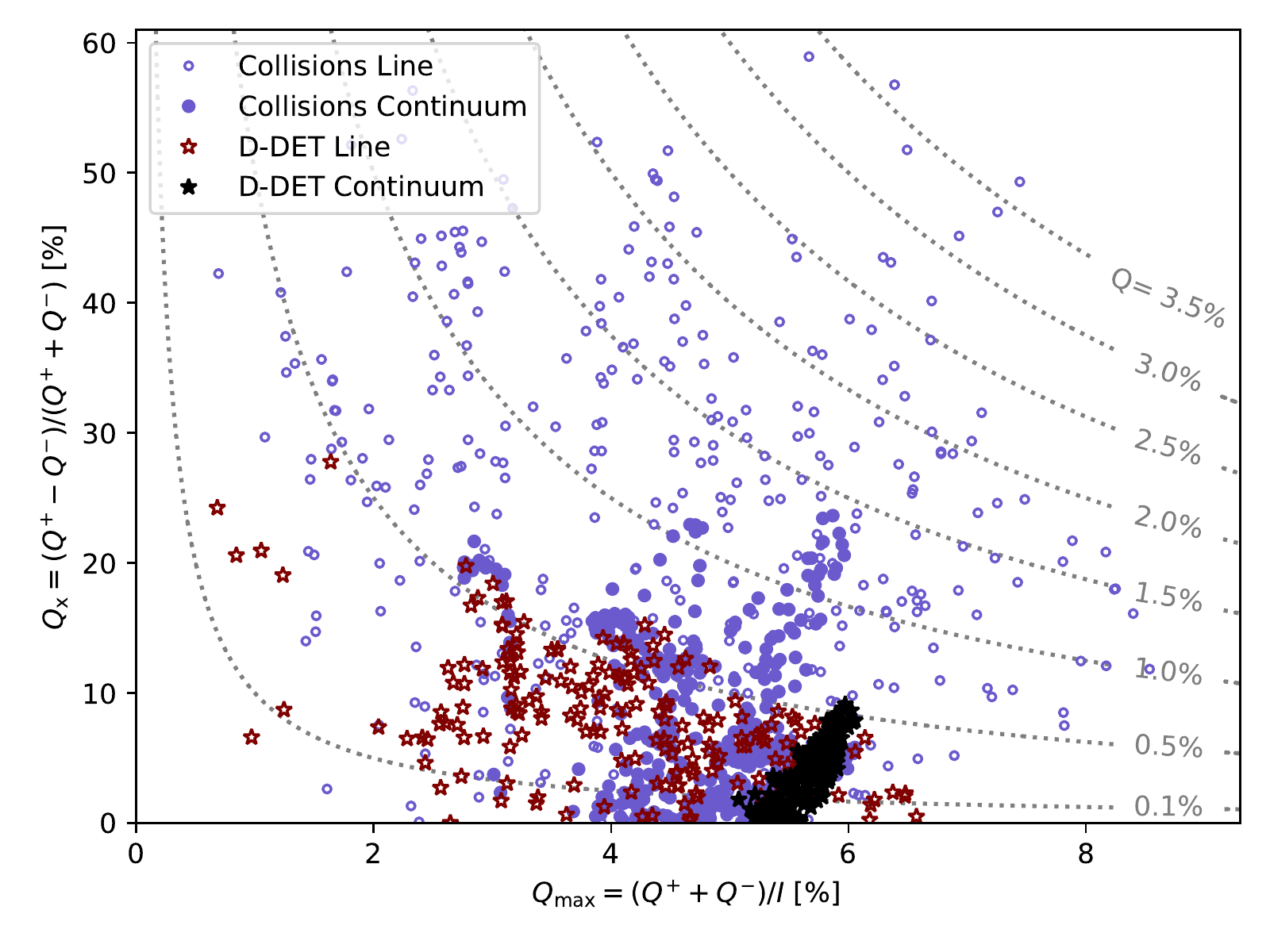}
	\caption{Breakdown of total polarization into $Q_{\max}$ and $Q_{\rm{x}}$ components (see Section~\ref{sec:Analysis}). The dotted lines represent contours of total $Q$ polarization. Blue filled and empty circles represent continuum and maximum \ion{Si}{II} line polarizations of 2D head-on collision models. Filled and empty stars represent continuum and maximum \ion{Si}{II} line polarization of double-detonation models (see Section~\ref{sec:Model}). All models are simulated with $v_{\rm{ph}}=10,000$~km/s. Each point represents one viewing angle. Note that the line polarization values on this plot are obtained without subtracting the continuum polarization and are therefore not identical to those presented in Fig.~\ref{fig:max_vs_ni56}.}
	
	\centering
	\label{fig:Breakdown}
\end{figure}

As detailed below, we find that $Q_{\max}$ depends on the optical depth for electron scattering in the ejecta. Fig.~\ref{fig:Qmax} shows this relation for several hydrodynamical and synthetic models. Remarkably, $Q_{\max} \sim 5\%$ universally for all the simulated models with a typical electron scattering optical depth of order unity $\tau_{\rm{es}} \sim 1$. Fig.~\ref{fig:Breakdown} plots $Q_{\max}$ and $Q_{\rm{x}}$ values for the 2D head-on collision and double-detonation models, highlighting the similarity in $Q_{\max}$ as opposed to the difference in the distribution of $Q_{\rm{x}}$, and accounting for the difference in the predicted total polarization of these models.

\subsection{Maximum Polarization - $Q_{\max}$}
\label{sec:Qmax}

Let us consider for simplicity the emission of polarized light in a spherically symmetric configuration. Polarization due to electron scattering is proportional to $1-\cos^2\chi$ (where $\chi$ is the scattering angle). Much of the light emitted from the ejecta comes from close to the center of the projected disk. For $\tau_{\rm{es}} \sim 1$, this is a mix of unpolarized light emitted from the photosphere and scattered light with $\chi$ close to $\pi$, and thus has low polarization. 

In \citet{Kasen2003}, the polarization of light emitted by simple electron-scattering atmospheres with a power-law electron density $\rho\propto r^{-n}$ was computed using a Monte Carlo simulation. The results showed that as the impact parameter of the beams emerging from the ejecta grows, the intensity decreases, while the polarization increases. The combined effect is that contributions to the total polarization come mainly from a distinct ring with a radius close to that of the photosphere, depending on the optical depth of the electron-scattering atmosphere (see Fig.~\ref{fig:Kasen} in the Appendix). 
Thus we expect the value of $Q_{\max}$ to be related to the radial density profile of free electrons in the ejecta. 

In Fig.~\ref{fig:Qmax}, we investigate the dependence of (continuum) $Q_{\max}$ on the optical depth. We compare average $Q_{\max}$ values for the 2D head-on collision and double-detonation models with several synthetic models as a function of average electron scattering optical depth:
\begin{equation}
	\tau_{\rm{es}}=\frac{1}{N}\sum_{i=1}^{N}t_{\rm{exp}}\sigma_{\rm{T}}\int_{v_{\rm{ph}}} ^{\infty}n_{\rm{e},\it{i}}(v)dv
	\label{eq:tau_es}
\end{equation}
where $N$ is the number of viewing angles, $t_{\rm{exp}}$ is the time since the explosion and $n_{\rm{e},\it{i}}(v)$ is the electron number density for the \mbox{$i$-th} viewing angle at velocity $v$ at time $t_{\rm{exp}}$ (see Fig.~\ref{fig:ColModels_tau}). We find that the 2D head-on collision simulations behave similar to the power-law model, rising as the optical depth increases, while models with a discrete thin shell display a peak of $Q_{\max}$ which then drops as the optical depth increases. 

An analytic estimation of the polarization of an axis-symmetric nebula illuminated by a central point source was obtained in \citet{BrownMclean77} and \citet{BrownMcleanEmslie78}, where the polarization was found to be proportional to the angle-averaged electron-scattering optical depth, along with a shape correction factor and a factor depending on the viewing angle (see also \citealt{DessartHillier2011}). However, this result is valid only for optically thin nebula with $\tau_{\rm{es}} \ll 1$, whereas the typical optical depth for Thomson scattering in SNe Ia ejecta is of order unity:
\begin{equation}
	\tau_{\rm{es}}\sim \frac{M/(28~m_p)}{4\pi(vt)^2}\sigma_T \approx 1.1\times \frac{M}{M_{\odot}}\left(\frac{v}{10^4~\rm{km/s}}\right)^{-2}\left(\frac{t}{16 \rm{d}}\right)^{-2}
	\label{eq:tau}
\end{equation}

As can be seen in Fig.~\ref{fig:Qmax}, the optical depth is of order unity and $Q_{\max}\sim 5\%$ for all models. These results imply that even before any asymmetry is introduced, the maximum possible polarization of SNe Ia is of the order of $5\%$ due to radial structure considerations alone. 


\subsection{Polarization Cancellation - $Q_{\rm{x}}$}
\label{sec:Qx}

\begin{figure}
	\centering
	\includegraphics[clip,  width=\columnwidth]{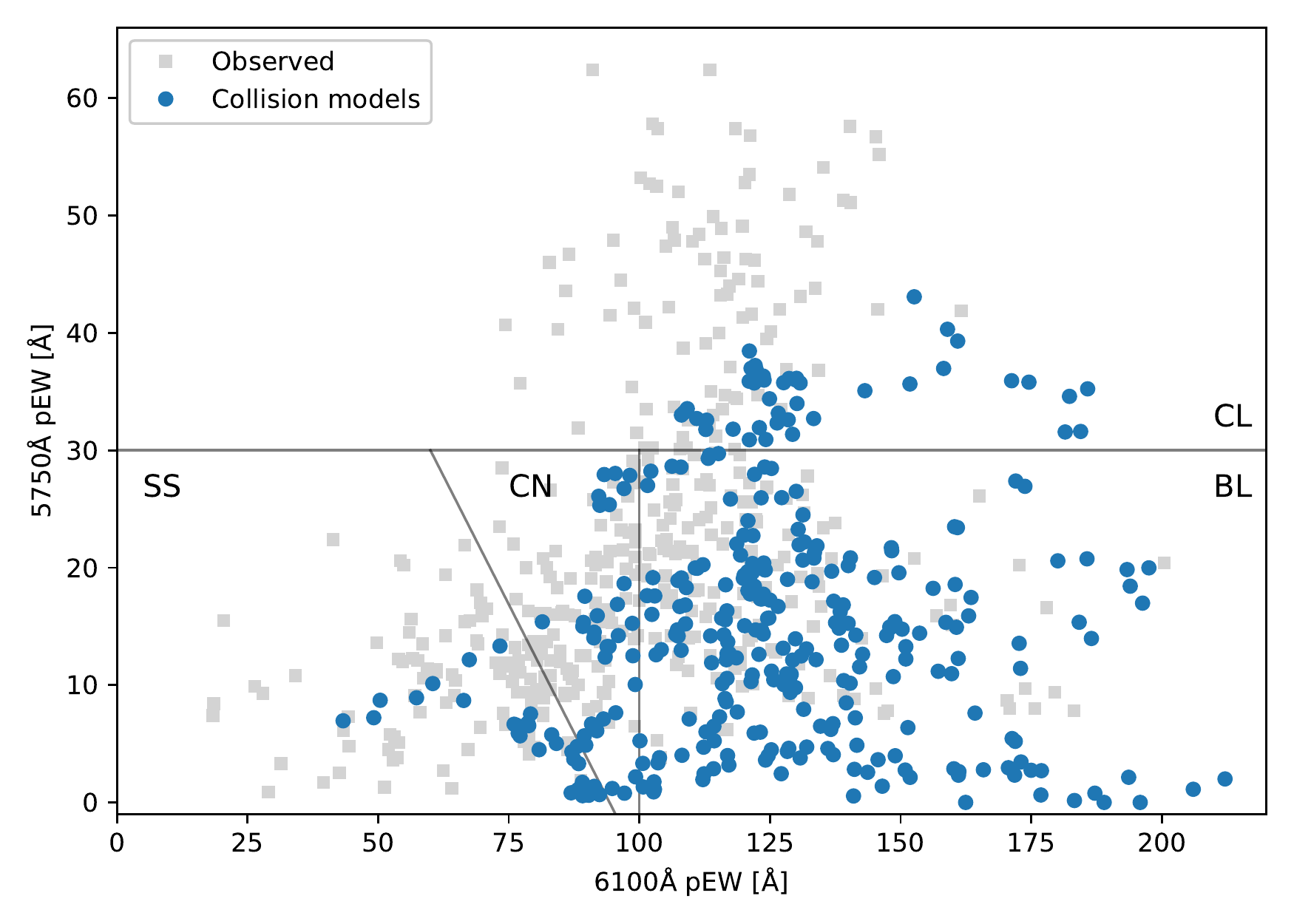}
	
	\caption{Branch plot of simulated 2D head-on collision models overlaid on observed data. Blue dots represent the pEWs extracted for different viewing angles of all head-on collision models with $v_{\rm{ph}}=10,000$~km/s, using the new code described in Section~\ref{sec:Methods}. Gray squares are observations, taken from \citet{CfA2012, CSP2013, Silverman2012}.}
	
	\centering
	\label{fig:Branch}
\end{figure}


The $Q_{\rm{x}}=\frac{Q^{+}-Q^{-}}{Q^{+}+Q^{-}}$ component represents the effect of cancellation on the total polarization. This factor depends on the spatial configuration of both the free electrons in the ejecta as well as the scattering elements which partially block the emerging light and depolarize it selectively. $Q_{\rm{x}}$ will be small for very symmetric ejecta ($Q^{+}\sim Q^{-}$) and can reach $100\%$ for a maximally asymmetric configuration, for example if $Q^{-}=0$. Note that $Q_{\rm{x}}$ can also be small for geometrically asymmetric ejecta, as long as the positive and negative contributions to $Q$ cancel out. 

In Fig.~\ref{fig:Breakdown} we plot the distribution of $Q_{\max}$ and $Q_{\rm{x}}$ values for the 2D head-on collision models. Here we again see that the collision models lie in a limited $Q_{\max}$ range around $\sim 5\%$ for both continuum and line polarization, noting that the $Q_{\max}$ range for line polarization is wider and can reach up to $8\%$. On the other hand, $Q_{\rm{x}}$ is quite variable: for continuum spectral regions $Q_{\rm{x}}\sim 10\%$, keeping the total polarization around $Q\sim0.5\%$. For the line spectral regions, we see a larger distribution of up to $Q_{\rm{x}}\sim50\%$, allowing the total polarization to extend up to $Q\sim3\%$. Note that the line polarization values on this plot are obtained without subtracting the continuum polarization and are therefore not identical to those presented in Fig.~\ref{fig:max_vs_ni56}.

Fig.~\ref{fig:Breakdown} also shows the continuum and line polarization decomposition for the double-detonation model (see Section~\ref{sec:Model}). It is interesting to note that this intrinsically asymmetric model spans a similar range of $Q_{\max}$, however, its line polarization is characterized by comparatively low $Q_{\rm{x}}$ and consequently low maximum line polarization of $Q\lesssim0.7\%$ for all viewing angles and all target luminosities. In continuum spectral regions we find a limit of $Q_{\rm{x}} \lesssim 10\% $, keeping the total $Q$ polarization lower than $\sim0.5\%$, similar to the collision model. 

\section{Branch plot}
\label{sec:BranchPlot}

\begin{figure}
	\centering
	\includegraphics[clip,  width=\columnwidth]{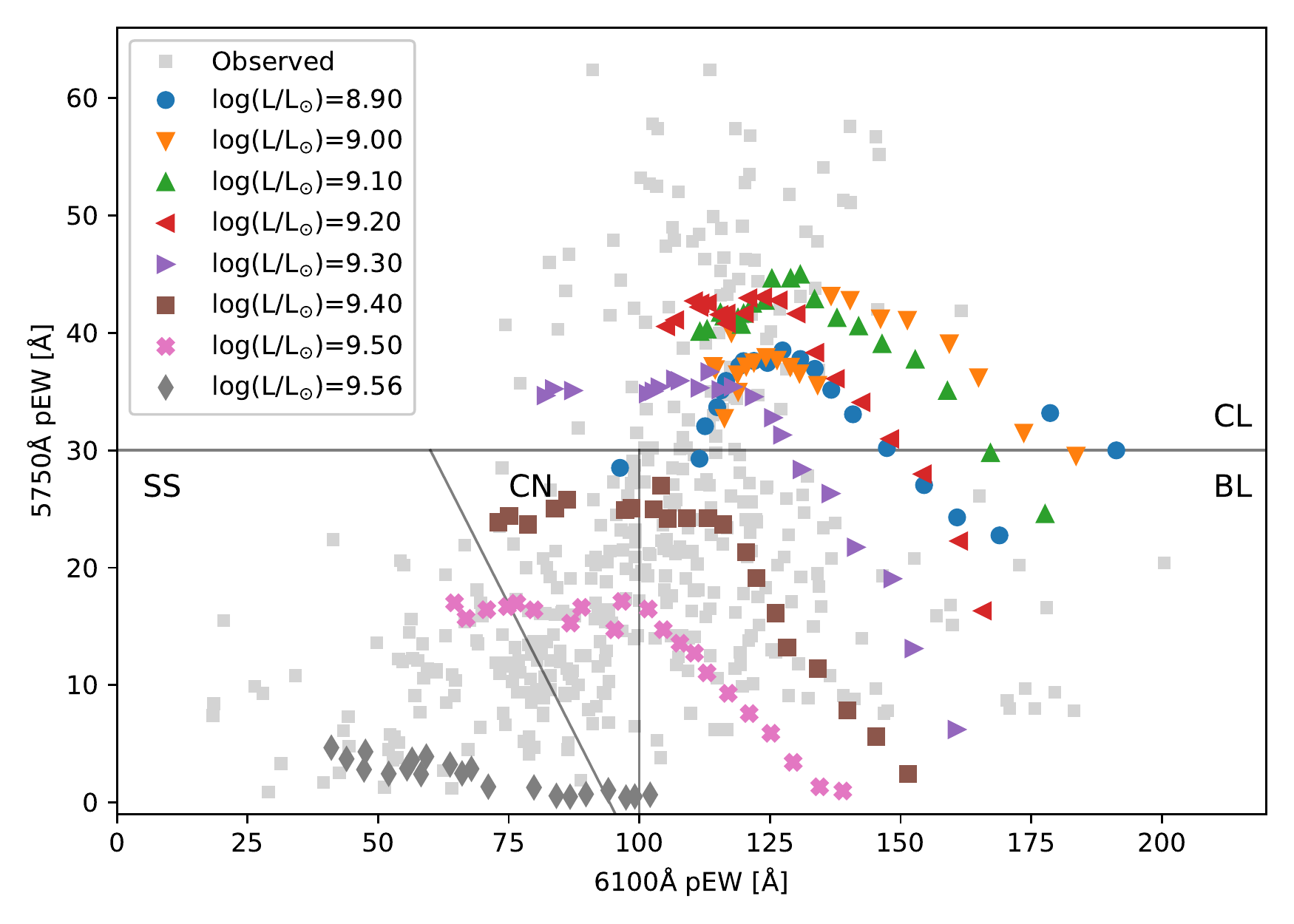}
	
	\caption{Branch plot of simulated double-detonation models overlaid on observed data. The model from \citet{Townsley2019} is synthetically extended (see Section~\ref{sec:Model}) - different markers represent different target luminosities. Each data point represents one viewing angle. Gray squares are observations, taken from \citet{CfA2012, CSP2013, Silverman2012}.}
	
	\centering
	\label{fig:Branch_Townsley}
\end{figure}


In \citet{Livneh2020}, we showed that the asymmetry of the head-on collision model allows it to span the two-dimensional distribution of \ion{Si}{II} line depths (Branch plot). In this paper, we use similar methods to calculate the electron densities and the optical depths of the various lines in the ejecta, but the final spectrum is obtained using the new 3D simulator described in Section~\ref{sec:Polarization}, in contrast to the 1D runs in the previous paper. For consistency, we verify that the resulting spectra still span the Branch plot. This is shown in Fig.~\ref{fig:Branch}.

Having found in Section~\ref{sec:ResultsLinePolarization} that the line polarization of the double-detonation model is low and consistent with the bulk of observations, we test how well it can cover the Branch plot. The results are shown in Fig.~\ref{fig:Branch_Townsley}: we find that it is able to span the Branch plot just as well as the head-on collision models. We note again that the models we use consist of a single hydro-dynamically simulated ejecta, artificially extended by performing the radiative transfer simulation using a range of target luminosities. Unfortunately, a full investigation of a range of hydro-dynamically simulated double-detonation models is beyond the scope of this paper. 


\section{Conclusions and Summary}

Polarization is an important indicator for SNe Ia, carrying information regarding the geometry of their ejecta which may be used to place constraints on their long-sought-after progenitor system/s and explosion mechanism/s. Observed SNe Ia polarization levels tend to be quite low both for continuum and line polarization, sometimes taken to suggest a high degree of symmetry in the ejecta \citep[e.g.][]{LivioMazzali2018}. On the other hand, many observed characteristics of SNe Ia, such as the variation in Si velocity gradients, bi-modal and shifted nebular spectral lines and the presence of high-velocity features, hint to significant asymmetry in the ejecta of these explosions \citep[e.g.][]{Maeda2010, Blondin2011, Childress2014, Kushnir2015, Dong2018, Maguire2018}.

In \citet[]{Livneh2020}, we showed that the distribution of \ion{Si}{II} pEWs (Branch plot) is naturally obtained in the asymmetric 2D ejecta produced by (zero impact parameter) head-on collisions of carbon-oxygen WDs. Radiation transfer was calculated in a simplified manner using 1D \textsc{Tardis} simulations along different lines of sight with a spherical photosphere. Despite the use of these rough approximations, the source of the wide distribution of \ion{Si}{II} pEWs was robustly related to the large variations in \ion{Si}{II} profiles along different lines of sight, suggesting that significant asymmetry in the \ion{Si}{II} profiles is required by the observations. 

Is such asymmetry consistent with polarization observations? 
To address this question we study the polarization signature of asymmetric explosion models using similar radiation transfer tools combined with a new 3D Monte Carlo polarization code (see Section~\ref{sec:Methods}). 
We focus on the same 2D collision model ejecta from \citet[]{Kushnir2013} that were studied in \citet[]{Livneh2020}, in addition to 2D asymmetric double-detonation ejecta from \citet{Townsley2019}. The latter are extended to represent a wide range of $^{56}$Ni yields by artificially using a range of luminosities in the calculations. 

In Section~\ref{sec:Analysis}, we show that the polarization $Q$ can be parametrized as a product $Q=Q_{\max}Q_{\rm{x}}$ (Eq.~\ref{eq:breakdown}) of a radial structure component $Q_{\max}$ which is insensitive to the models and a component $Q_{\rm{x}}$ which depends on the cancellation between opposite polarizations. The radial component is shown to be universally around $Q_{\max}\sim 5\%$ for SNe Ia models which have a typical electron scattering optical depth of order unity $\tau_{\rm{es}}\sim 1$, including the collision and the double-detonation models (see Fig.~\ref{fig:Qmax}). 
This result elucidates the simple reason why SNe Ia typically present such low polarization signatures: even before any asymmetry is introduced, the maximum possible polarization of SNe Ia is of the order of $5\%$ due to radial structure considerations alone.  

Generally speaking, the polarization of SNe Ia radiation arises from a combination of three sources: (i) asymmetry in the cloud of scattering electrons in the ejecta; (ii) asymmetry in the absorption following the scattering (focusing here on \ion{Si}{II} lines); and (iii) asymmetry in the distribution of emission before the scattering (mainly set by the iron group elements).
While simplified, the rough radiation transfer treatment that we use allows us to estimate the approximate scale of polarization associated with the first and second sources. 
A more detailed calculation that accounts for the iron-group element distribution is needed to verify that there is no accidental cancellation that arises. 

The asymmetry in the electron distribution is found to be low in both the collision and the double-detonation models, with continuum polarizations of $Q\sim 0.5\%$ (corresponding to cancellations of $Q_{\rm{x}}\sim 10\%$, see Figs.~\ref{fig:cont_vs_ni56_vph},~\ref{fig:Breakdown}). 
However, the \ion{Si}{II} line polarization signatures differ between the two models: the collision model results in high line polarizations reaching $\sim 3\%$ ($Q_{\rm{x}}\lesssim 50\%$, see Figs.~\ref{fig:max_vs_ni56},~\ref{fig:Breakdown}), in tension with observations (mostly $\lesssim 1.2\%$).  
In contrast, the double-detonation model results in low line polarizations $\lesssim 0.7\%$ ($Q_{\rm{x}}\sim 10\%$, see Figs.~\ref{fig:max_vs_ni56_vph},~\ref{fig:Breakdown}) consistent with previous studies of double-detonation models \citep{Bulla2016.2} and with most observations. 

The simulated polarization's sensitivity to the choice of photospheric velocity $v_{\rm{ph}}$ is explored for the collision model in Section~\ref{sec:ResultsLinePolarization}. Choosing a higher $v_{\rm{ph}}$ would in most cases bring down the polarization levels. However, this would result in inconsistent \ion{Si}{II} velocities, and thus a photosphere with high velocity is not a promising way to resolve the tension with observations. 
On the other hand, we note that the zero impact parameter models represent an extreme case of WD collisions, perhaps with extreme polarization, whereas 3D simulations may produce different results. Also, the limited sample of SNe Ia spectral polarization observations may lead to biases: several additional high-polarization events may significantly reduce the tension.

In Section~\ref{sec:BranchPlot}, we show that the asymmetric \ion{Si}{II} distribution in the ejecta of both models is able to reproduce the observed Branch plot (see Figs.~\ref{fig:Branch},~\ref{fig:Branch_Townsley}). 
However, whereas the 2D head-on collision model's line polarization is too high to be consistent with most observations, the double-detonation model provides an example of a significantly asymmetric explosion, sufficient to explain the distribution of \ion{Si}{II} pEWs in the Branch plot, while simultaneously reproducing the observed low levels of \ion{Si}{II} line polarization. 
An examination of Fig.~\ref{fig:scatters} provides qualitative insight as to the difference between the two models: the irregular Si distribution of the shown collision model leads to viewing angles in which the \ion{Si}{II} ions block (or depolarize) polarized light from predominantly $Q^+$ or $Q^-$ regions, whereas the spheroid-like structure of the double-detonation model ejecta (see Fig.~\ref{fig:Townsley_ejecta}) tends to leave similar contributions from both $Q^+$ and $Q^-$ regions, resulting in a large degree of cancellation and consequently low values of $Q_{\rm{x}}$.

These results demonstrate that there is no inherent contradiction between the required asymmetry and a low polarization signature, strengthening the case for asymmetric explosions as progenitors of SNe Ia. 
The high polarization obtained for the 2D collision model ejecta highlights the need for large cancellations to account for the low observed line polarizations in such ejecta. The combination of significant asymmetry, able to account for the Branch plot, together with sufficient regularity to be consistent with low polarization, seems to be a stringent combination of constraints on models of SNe Ia explosions.
Our calculations suggest that these constraints are satisfied by the double-detonation model, whereas we find no evidence that the 2D head-on collision models can produce cancellations that result in low \ion{Si}{II} polarization signatures. However, our results are obtained within the context of limited 2D simulations, which may not capture all the relevant physics. Whether or not other asymmetric models such as a more realistic 3D ejecta of WD collisions with non-zero impact parameters can pass this high bar is yet to be seen.


\section*{Acknowledgements}

We thank Doron Kushnir and Dean Townsley for sharing model ejecta.
We thank Nahliel Wygoda for sharing light-curve data.
We thank Eli Waxman, Avishay Gal-Yam, Doron Kushnir and Subo Dong for useful discussions.
We thank the anonymous referee for helpful comments.
This work was supported by the Beracha Foundation and the Minerva foundation with funding from the Federal German Ministry for Education and Research.
This research made use of \textsc{Tardis}, a community-developed software package for spectral synthesis in supernovae \added{\mbox{\citep{Tardis, kerzendorf_wolfgang_2019_2590539}}}. The development of \textsc{Tardis} received support from the Google Summer of Code initiative and ESA's Summer of Code in Space program. \textsc{Tardis} makes extensive use of Astropy and PyNE.


\bibliographystyle{mnras}
\bibliography{Pol} 

\clearpage 

\appendix


\section{Polarization Simulator Verification}
\label{sec:SimulatorVerification}

In order to verify our new polarization simulator (described in Section~\ref{sec:Polarization}), we compare our results with several previous works.


\subsection{Flux Spectrum}

\begin{figure}
	\centering
	\includegraphics[clip,  width=\columnwidth]{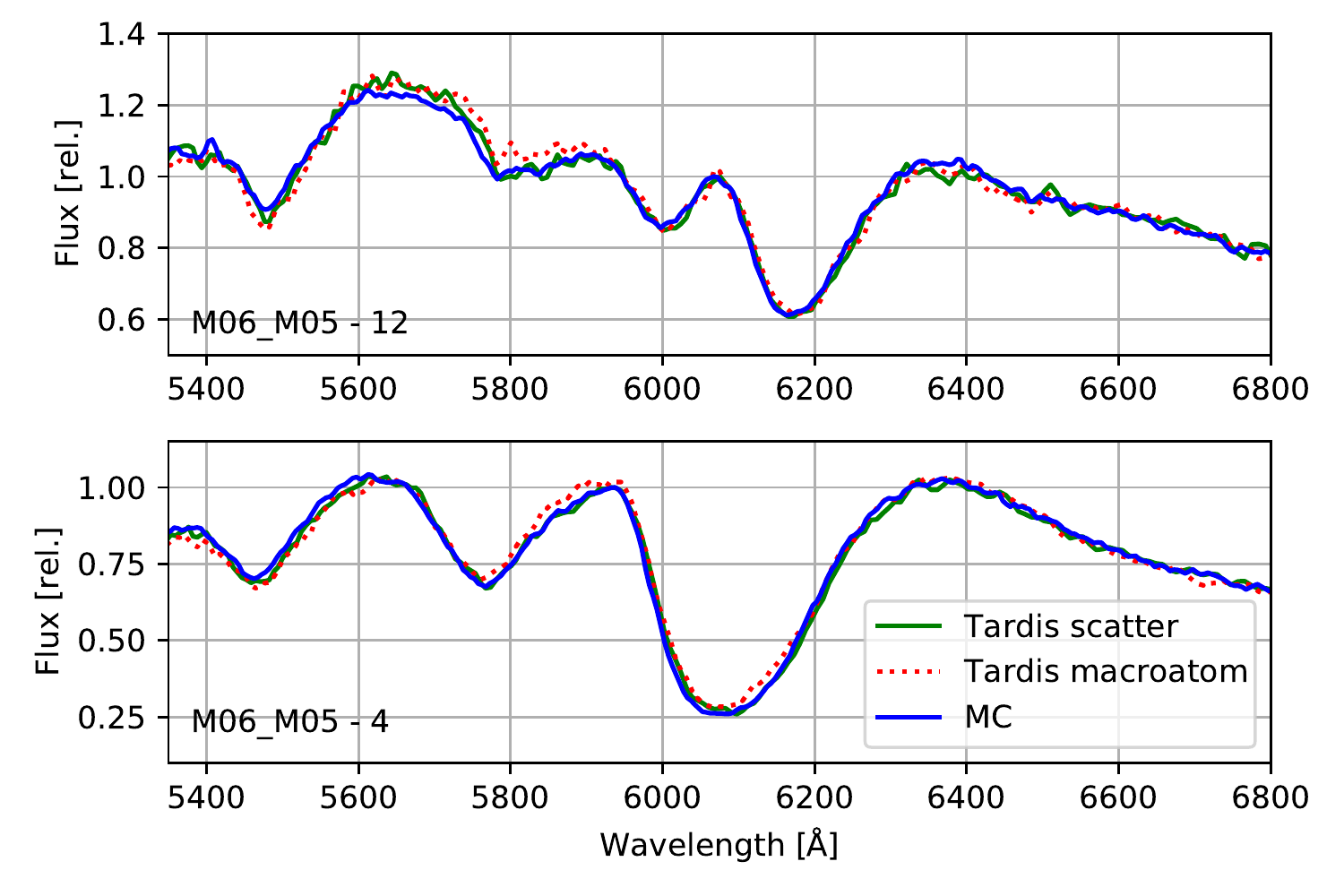}
	\caption{Flux spectrum for two example viewing angles of the collision model \texttt{M06\_M05}. The spectrum is computed with \textsc{Tardis} using two line-interaction modes (\texttt{scatter} and \texttt{macroatom}) and compared to the results of our Monte Carlo simulator.}
	
	\centering
	\label{fig:Tardis}
\end{figure}

First, we verify that the flux spectrum obtained from our simulator is identical to the flux spectrum given by \textsc{Tardis}. As an example, we take two viewing angles from the head-on collision model \texttt{M06\_M05}. We simulate radiative transfer with \textsc{Tardis} and record the resulting electron densities, optical depths for 35 absorption lines with $\tau>0.1$ and the photospheric temperature. Next, we run the MC simulator with these parameters, taking all viewing angles to be identical in order to emulate the 1-D \textsc{Tardis} setup. Finally, we compare the flux spectrum computed by our MC simulator to the \textsc{Tardis} spectrum. The results are shown in Fig.~\ref{fig:Tardis}. We note that for this comparison, the \textsc{Tardis} line interaction mode was chosen to be \texttt{scatter}, since our simulator does not support \texttt{macroatom}. Both interaction modes are shown for comparison. 


\subsection{Continuum Polarization}
\label{sec:SimulatorVerificationContinuum}

\begin{figure}
	\centering
	\includegraphics[clip,  width=\columnwidth]{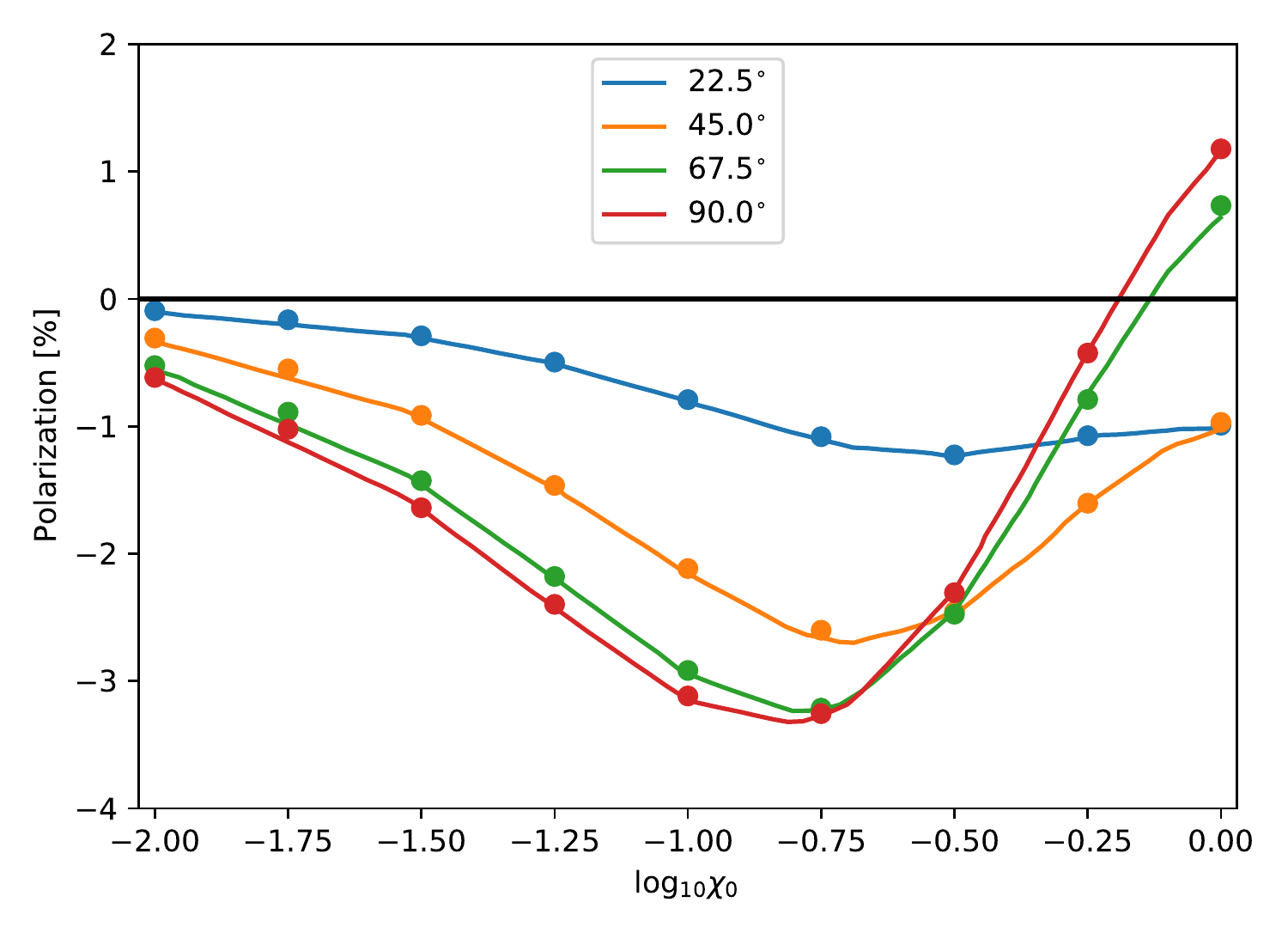}
	\caption{Continuum polarization as a function of $\chi_0$ for different viewing angles, for the model described in Section~\ref{sec:SimulatorVerificationContinuum}. Lines are reproduced from \citet{Hillier1994}, while dots show results obtained using our polarization simulator.}
	\centering
	\label{fig:Hillier}
\end{figure}

Next we verify the correct treatment of the polarization. We simulate the following model from \citet{Hillier1994}: A point source is surrounded by a detached spherical shell with inner radius $R_{\min}=2.0$ and outer radius $R_{\max}=30.0R_{\min}$ with a prolate electron number density distribution $N_{\rm{e}}\left(r,\beta\right)$ such that
\begin{equation}
	\sigma_{\rm{e}}N_{\rm{e}}\left(r,\beta\right)=\chi_0\left(\frac{R_{\min}}{r}\right)^4\left(1+10{\cos}^2{\beta}\right)
\end{equation}
where $r$ and $\beta$ express the radius and the polar angle inside the envelope. The resulting polarization as a function of viewing angle (Fig.~\ref{fig:Hillier}) agree with the results given in \citet[][Fig. 2]{Hillier1994}. 


\subsection{Impact Parameter}
\begin{figure}
	\centering
	\includegraphics[clip,  width=\columnwidth]{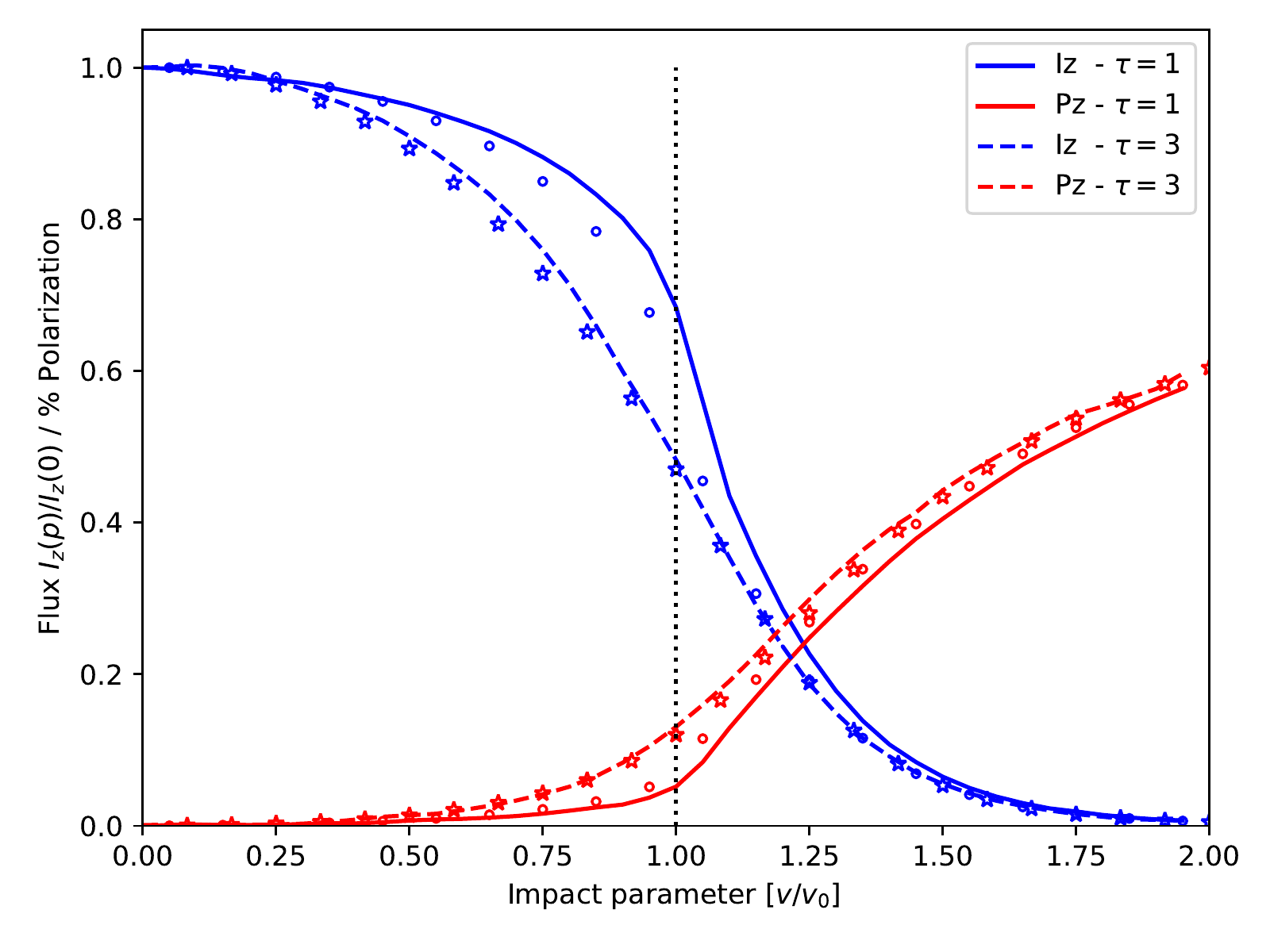}
	
	\caption{Intensity (Iz) and polarization (Pz) of beams emerging from an electron scattering ejecta as a function of the impact parameter for a synthetic model from \citet{Kasen2003} with a density profile of $\rho\propto r^{-7}$. The optical depth from the inner boundary to infinity is set at $\tau_{\rm{es}}=3$ and $\tau_{\rm{es}}=1$. The impact parameter is given in units of the photospheric velocity $v_0$, defined as continuum optical depth of 1. Lines are results from \citet[][Fig. 5]{Kasen2003}, stars and circles are results from our polarization simulator.}
	
	\centering
	\label{fig:Kasen}
\end{figure}

Another interesting comparison is the intensity and polarization of beams emerging from an ejecta as a function of the impact parameter. We simulate a synthetic model introduced in \citet{Kasen2003}: an inner boundary surface, surrounded by an electron-scattering envelope with a power-law electron density $\rho\propto r^{-7}$. The optical depth from the inner boundary to infinity is set at $\tau_{\rm{es}}=3$ and $\tau_{\rm{es}}=1$. Results are shown in Fig.~\ref{fig:Kasen}. We find that our results agree qualitatively with a small discrepancy, which may be due to binning and other variations in implementation.


\label{lastpage}
\end{document}